\documentclass[12pt]{article}

\usepackage{scicite}

\usepackage{graphics}
\usepackage{color}
\usepackage{float}
\usepackage{graphicx}
\usepackage{amssymb}
\usepackage{overpic}
\usepackage{epstopdf}
\usepackage{wasysym}
\usepackage{bm}
\usepackage{graphicx}

\newenvironment{sciabstract}{%
\begin{quote} \bf}
{\end{quote}}

\newenvironment{methods}{%
    \section*{Methods}%
    \setlength{\parskip}{0pt}%
    }{}

\makeatletter
\def\@cite#1#2{\textsuperscript{{#1\if@tempswa , #2\fi}}}
\makeatother

\newcommand{\onlinecite}[1]{\hspace{-1 ex} \nocite{#1}\citenum{#1}}

\newcommand{\lesco}{La$_{1.7}$Eu$_{0.2}$Sr$_{0.1}$CuO$_{4}$}
\newcommand{\lnsco}{La$_{1.48}$Nd$_{0.4}$Sr$_{0.12}$CuO$_{4}$}
\newcommand{\lsco}{La$_{2-x}$Sr$_{x}$CuO$_{4}$}
\newcommand{\lbco}{La$_{2-x}$Ba$_x$CuO$_4$}

\newcounter{lastnote}

\renewcommand{\figurename}{{\bf{Figure}}}

\makeatletter
\makeatletter \renewcommand{\fnum@figure}{{\bf{\figurename~\thefigure}}}
\makeatother

\title{Signatures of a pair density wave at high magnetic fields in cuprates with charge and spin orders}

\author
{Zhenzhong Shi,$^{1\dag}$ P. G. Baity,$^{1,2\dag\dag}$ J. Terzic,$^{1}$ T. Sasagawa,$^{3}$ Dragana Popovi\'{c}$^{1,2\ast}$\\
\\
\normalsize{$^{1}$National High Magnetic Field Laboratory, Florida State University,}\\
\normalsize{Tallahassee, Florida 32310, USA}\\
\normalsize{$^{2}$Department of Physics, Florida State University,}\\
\normalsize{Tallahassee, Florida 32306, USA}\\
\normalsize{$^{3}$Materials and Structures Laboratory, Tokyo Institute of Technology,}\\
\normalsize{Kanagawa 226-8503, Japan}\\
\\
\normalsize{$^{\dag}$ Present address: Department of Physics, Duke University,}\\
\normalsize{Durham, North Carolina 27708, USA}\\
\normalsize{$^{\dag\dag}$ Present address: James Watt School of Engineering, University of Glasgow,}\\
\normalsize{Glasgow, G12 8QQ, Scotland, United Kingdom}\\
\normalsize{$^\ast$To whom correspondence should be addressed; E-mail: dragana@magnet.fsu.edu}
}

\date{}

\begin{document} 


\baselineskip24pt


\maketitle 


\begin{sciabstract}
In underdoped cuprates, the interplay of the pseudogap, superconductivity, and charge and spin ordering can give rise to exotic quantum states, including the pair density wave (PDW), in which the superconducting (SC) order parameter is oscillatory in space.  However, the evidence for a PDW state remains inconclusive and its broader relevance to cuprate physics is an open question.  To test the interlayer frustration, the crucial component of the PDW picture, we performed transport measurements on La$\bm{_{1.7}}$Eu$\bm{_{0.2}}$Sr$\bm{_{0.1}}$CuO$\bm{_{4}}$ and La$\bm{_{1.48}}$Nd$\bm{_{0.4}}$Sr$\bm{_{0.12}}$CuO$\bm{_{4}}$, cuprates with ``striped'' spin and charge orders, in perpendicular magnetic fields ($\bm{H_\perp}$), and also with an additional field applied parallel to CuO$_2$ layers ($\bm{H_\parallel}$).  We detected several phenomena predicted to arise from the existence of a PDW, including an enhancement of interlayer SC phase coherence with increasing $\bm{H_\parallel}$.  Our findings are consistent with the presence of local, PDW pairing correlations that compete with the uniform SC order at $\bm{T_{c}^{0}< T<(2-6) T_{c}^{0}}$, where $\bm{T_{c}^{0}}$ is the $\bm{H=0}$ SC transition temperature, and become dominant at intermediate $\bm{H_\perp}$ as $\bm{T\rightarrow 0}$.  These data also provide much-needed transport signatures of the PDW in the regime where superconductivity is destroyed by quantum phase fluctuations.
\end{sciabstract}

The origin of the cuprate pseudogap regime has been a long-standing mystery.  The richness of experimental observations\cite{Keimer2015} and the instability of underdoped cuprates towards a variety of ordering phenomena, such as periodic modulations of charge density discovered in all families of hole-doped cuprates\cite{Comin2016}, have raised the possibility that putative PDW correlations\cite{Himeda2002,Berg2009-PRB} may be responsible for the pseudogap regime\cite{Fradkin2015,Agterberg2019}.  In order to distinguish between different scenarios, the most intriguing open question is what happens at low $T\ll T_{c}^{0}$ and high $H_\perp$, when SC order is destroyed by quantum phase fluctuations\cite{Agterberg2019} and short-range charge orders are enhanced\cite{Wen2012,Huecker2013,Gerber2015}.  However, the experimental evidence for a PDW state remains scant and largely indirect in the first place.

A PDW SC state was proposed\cite{Berg2007,Berg2009-PRB} to explain the suppression of the interlayer ($c$-axis) Josephson coupling (or dynamical layer decoupling) apparent in the $H=0$ anisotropic transport\cite{Li2007} in La$_{1.875}$Ba$_{0.125}$CuO$_4$, as well as in optical measurements in La$_{1.85-y}$Nd$_y$Sr$_{0.15}$CuO$_4$ when the Nd concentration was tuned into the stripe-ordered regime\cite{Tajima2001}.  The dynamical layer decoupling was observed also  in the presence of an applied $H_\perp$, in La$_{1.905}$Ba$_{0.095}$CuO$_4$ (ref.~\onlinecite{Stegen2013}) and  \lsco\, (ref.~\onlinecite{SchafgansPRL2010}).  In La$_{2-x-y}$(Ba,Sr)$_x$(Nd,Eu)$_y$CuO$_4$ compounds near $x=1/8$, charge order appears in the form of stripes, which are separated by regions of oppositely phased antiferromagnetism (``spin stripes'')\cite{Fradkin2015}  at $T<T_{SO}<T_{CO}$; here $T_{SO}$ and $T_{CO}$ are the onsets of spin and charge stripes, respectively.  In \lsco\, at $x=0.10$, spin stripe order is induced\cite{Lake2002} by applying $H_\perp$.  The dynamical layer decoupling was thus attributed\cite{Berg2009-PRB,Berg2007} to a PDW SC state\cite{Berg2007,Himeda2002}, such that the spatially modulated SC order parameter, with zero mean, occurs most strongly within the charge stripes, but the phases between adjacent stripes are reversed (antiphase).  Since stripes are rotated by $90^{\circ}$ from one layer to next, antiphase superconductivity within a plane strongly frustrates the interlayer SC phase coherence\cite{Fradkin2015}, leading to an increase in anisotropy.  This effect is reduced by doping away from $x=1/8$, but $H_\perp$ can lead to dynamical layer decoupling as static stripe order is stabilized by a magnetic field.

To obtain more definitive evidence of the existence of a PDW, recent experiments have focused on testing various theoretical predictions\cite{Fradkin2015}.  
For example, transport measurements on La$_{1.875}$Ba$_{0.125}$CuO$_4$ have employed $H_\perp$ high enough to decouple the planes and then to suppress the SC order within the planes, with the results consistent with pair correlations surviving in charge stripes\cite{Li2019}; Josephson junction measurements\cite{Hamilton2018} on La$_{1.875}$Ba$_{0.125}$CuO$_4$ devices support the prediction of a charge-4$e$ SC condensate, consistent with the presence of a PDW state; an additional charge order was detected\cite{Edkins2018} in Bi$_2$Sr$_2$CaCu$_2$O$_8$ by scanning tunneling microscopy (STM) at very low $H_\perp/T_{c}^{0}\lesssim 0.1$~T/K, consistent with a PDW order that emerges within the halo region surrounding a vortex core once a uniform SC order is sufficiently suppressed by $H_\perp$.  However, alternative explanations are still possible, and additional experiments are thus needed to search for a PDW and explore its interplay with other orders in the pseudogap regime\cite{Agterberg2019}.

Therefore, we measured transport in La$_{2-x-y}$Sr$_x$(Nd,Eu)$_y$CuO$_4$ compounds, which have the same low-temperature structure as \lbco, over an unprecedented range of $T$ down to $T/T_{c}^{0} \lesssim 0.003$ and fields up to $H/T_{c}^{0}\sim 10$~T/K.  We combined linear in-plane resistivity $\rho_{ab}$, nonlinear in-plane transport or voltage-current ($V$--$I$) characteristics, and the anisotropy ratio $\rho_c/\rho_{ab}$ (here $\rho_c$ is the out-of-plane resistivity) to probe both charge and vortex matter on single crystals with the nominal composition \lesco\, and \lnsco\, (Methods); the former is away from $x=1/8$ and thus the stripe order is weaker\cite{Fradkin2015}.  We find signatures of dynamical layer decoupling in both $H=0$ and with increasing $H_\perp$, consistent with the presence of a PDW.  However, a key proposed test of this interpretation involves relieving the interlayer frustration through the application of an in-plane magnetic field\cite{Berg2007,Fradkin2015}.  In particular, since $H_\parallel$ can reorient the spin stripes in every other plane\cite{Huecker2004,Huecker2008,Baek2015}, a consequence of a PDW would be an enhancement of interplane coherence, or a reduced anisotropy.  This is precisely what we test and observe.

\begin{figure}[!tb]
\centerline{\includegraphics[width=\textwidth,clip=]{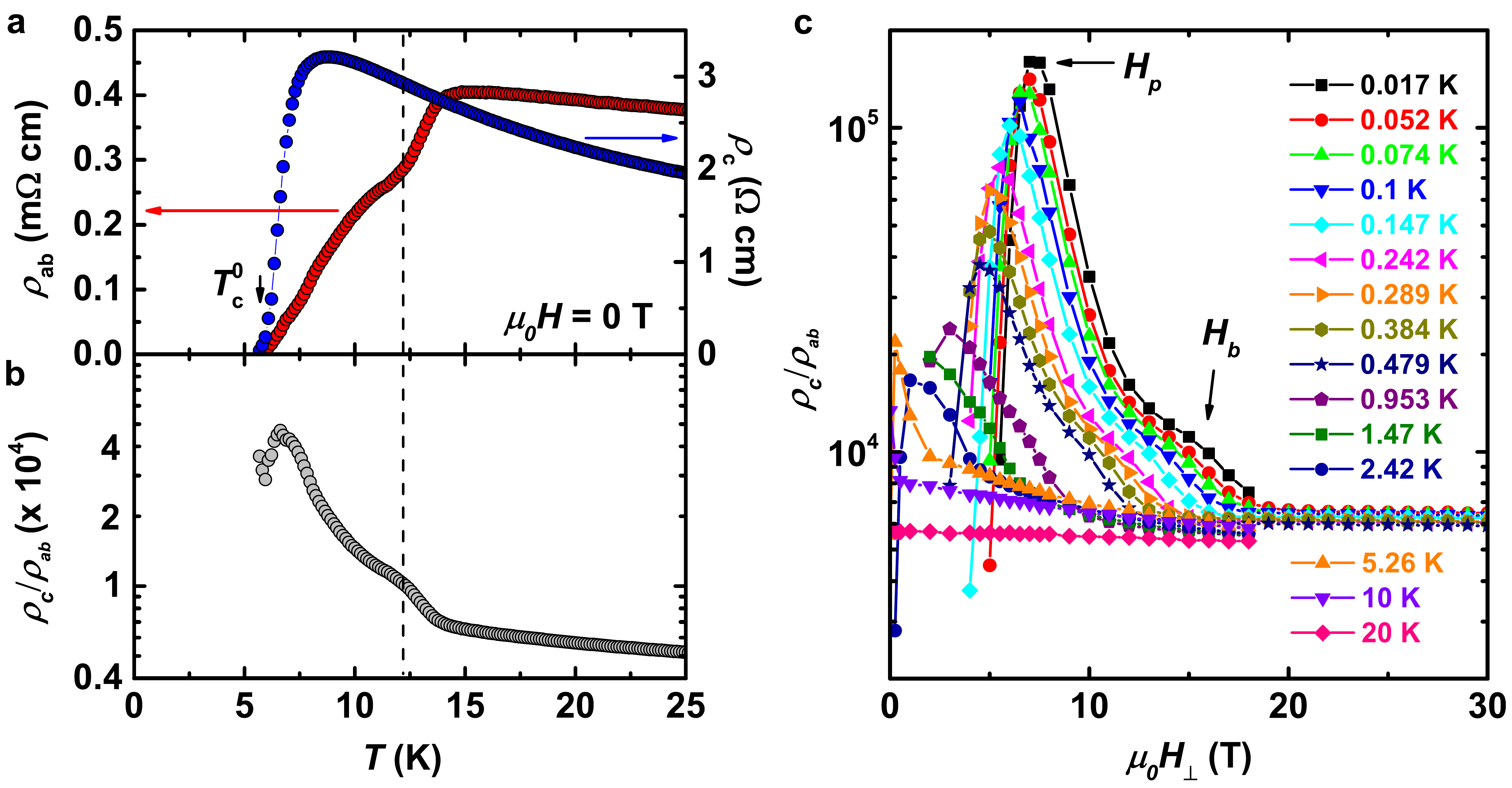}} 
\caption{\textbf{Evolution of the anisotropy in La$\bm{_{1.7}}$Eu$\bm{_{0.2}}$Sr$\bm{_{0.1}}$CuO$\bm{_4}$ with $\bm{T}$ and $\bm{H_\perp}$.}
\textbf{a}, $\rho_{ab}(T)$ and $\rho_c(T)$, and  \textbf{b}, the anisotropy ratio $\rho_c/\rho_{ab}(T)$, in zero field.  The vertical dashed line indicates where SC correlations are established in the planes, resulting in the enhancement of the anisotropy; $\rho_c$ continues to grow with decreasing $T$.  \textbf{c}, $\rho_c/\rho_{ab}$ vs $H_{\perp}$ at different $T$, as shown.  Arrows show the positions of the anisotropy peak $H_p$, or the decoupling field, as well as $H_b$, where the anisotropy is enhanced.  The method to determine $H_b$ more precisely is described in Supplementary Fig.~1.
} 
\label{new-anis1}
\end{figure}

We first explore the anisotropy in $H=0$.  In both \lesco\, and \linebreak \lnsco, $\rho_c$ and  $\rho_{ab}$ vanish at the same $T_{c}^{0}$ within the error (Methods; see also Supplementary Information), indicating the onset of 3D superconductivity, similar to \lsco\, (e.g. ref.~\onlinecite{Ando1995}).  The initial drop of $\rho_{ab}(T)$ with decreasing $T$ (Fig.~\ref{new-anis1}a) is accompanied by an enhancement of the anisotropy (Fig.~\ref{new-anis1}b), which continues to increase by almost an order of magnitude as $T$ is 
%
%
lowered further towards $T_{c}^{0}$.  These data look remarkably similar to those on La$_{1.875}$Ba$_{0.125}$CuO$_4$ (ref.~\onlinecite{Li2007}) that motivated theoretical proposals for a PDW SC state in striped cuprates: the initial, high-$T$ enhancement of the anisotropy is understood to reflect the establishment of SC correlations in CuO$_2$ planes.  

The evolution of $\rho_c/\rho_{ab}(T)$ with $H_\perp$ is shown in Fig.~\ref{new-anis1}c.   The anisotropy at the highest $T=20$~K is $\rho_c/\rho_{ab}\sim 6000$ and practically  independent of $H_\perp$.  However, as $T$ is lowered below $T_{c}^{0}$, $\rho_c/\rho_{ab}$ develops a distinctly nonmonotonic behavior as a function of $H_\perp$.  At $T=0.017$~K, for example, the anisotropy increases with $H_\perp$ by over an order of magnitude before reaching a peak ($\rho_c/\rho_{ab}>10^5$) at $H_\perp=H_p$, signifying decoupling of or the loss of phase coherence between the planes.  However, strong SC correlations persist in the planes for $H_\perp>H_p$: here $\rho_c/\rho_{ab}$ decreases with $H_\perp$ to $H_\perp$-independent values, comparable to those at high $T$, for the highest $H_\perp>20$~T.  This is in agreement with previous evidence\cite{ZShi2019-Hc2} that the $H_\perp>20$~T region corresponds to the normal state.  A smooth, rapid decrease of the anisotropy for $H_\perp>H_p$ is interrupted by a ``bump'' or an enhancement in $\rho_c/\rho_{ab}$, centered at $H_b$.  Therefore, the behavior of $\rho_c/\rho_{ab}$ is qualitatively the same whether the SC transition is approached from either (1) the high-$T$ normal state by lowering $T$ in $H=0$ (Fig.~\ref{new-anis1}b) or (2) the high-$H_\perp$ normal state by reducing $H_\perp$ at a fixed $T$ (Fig.~\ref{new-anis1}c).  These results thus suggest that the enhancement of the anisotropy near $H_b(T)$ may be attributed to the establishment of SC correlations in the planes as the SC transition is approached from the high-field normal state.

\begin{figure}[!tb]
\centerline{\includegraphics[width=\textwidth]{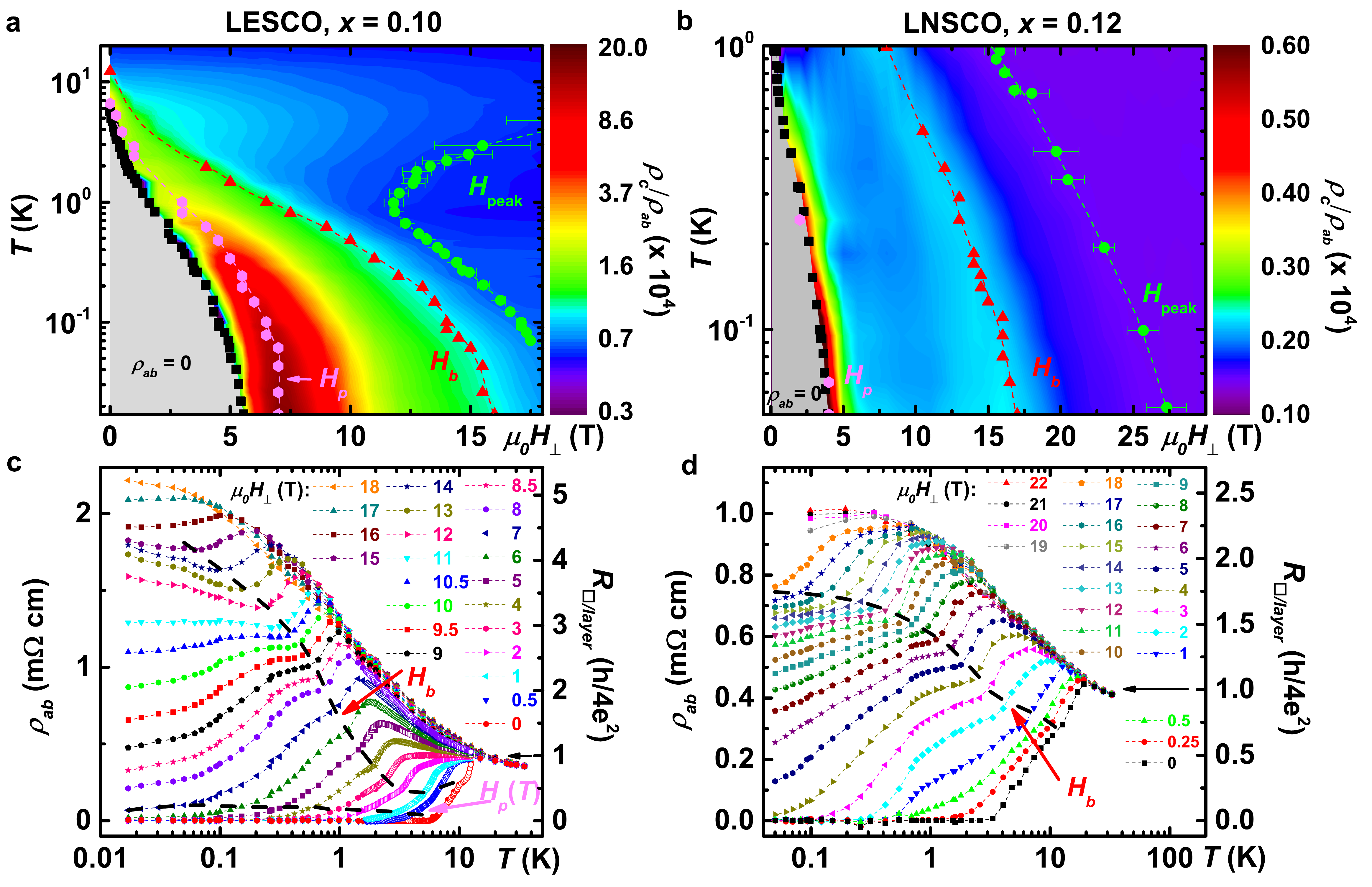}}
\caption{\textbf{Comparison of the anisotropy and the in-plane resistivity for different $\bm{T}$ and $\bm{H_\perp}$.} 
The color map in \textbf{a} and \textbf{b} shows $\rho_c/\rho_{ab}$ in La$_{1.8-x}$Eu$_{0.2}$Sr$_x$CuO$_4$ (LESCO) with $x=0.10$ (data from Fig.~\ref{new-anis1}c) and La$_{1.6-x}$Nd$_{0.4}$Sr$_x$CuO$_4$ (LNSCO) with $x=0.12$, respectively.  Black squares: $T_c(H_\perp)$; $\rho_{ab}=0$ for all $T<T_c(H_\perp)$.  Green dots: $H_{peak}(T)$, i.e. fields above which the in-plane MR changes from positive to negative; it has been established\cite{ZShi2019-Hc2} that $H_{peak}(T)\sim H_{c2}(T)$, i.e. the upper critical field.  Pink dots: $H_{p}(T)$, the layer decoupling field; red triangles: $H_{b}(T)$, where SC correlations are established in the planes as the SC transition is approached from the normal state.  $\rho_{ab}(T)$ of \textbf{c}, \lesco, and \textbf{d}, \lnsco\, for several $H_\perp$, as shown.   Open symbols in \textbf{c} show the data from another run.  Short-dashed lines guide the eye.  The $H_b(T)$ values obtained from the anisotropy are represented by the black dashed lines, as shown.  The lower black dashed line in \textbf{c} corresponds to the layer decoupling field, $H_{p}(T)$.  In \textbf{d}, $H_p(T)\gtrsim H_c(T)$ [or $T_c(H_\perp)$].  Black arrows in \textbf{c} and \textbf{d} show that the splitting of the $\rho_{ab}(T)$ curves for different $H_\perp$ becomes pronounced when $R_{\square/\mathrm{layer}}\approx R_{Q}=h/(2e)^2$. 
}
\label{new-anis2}
\end{figure}

\begin{figure}[!tb]
\centerline{\includegraphics[width=0.81\textwidth]{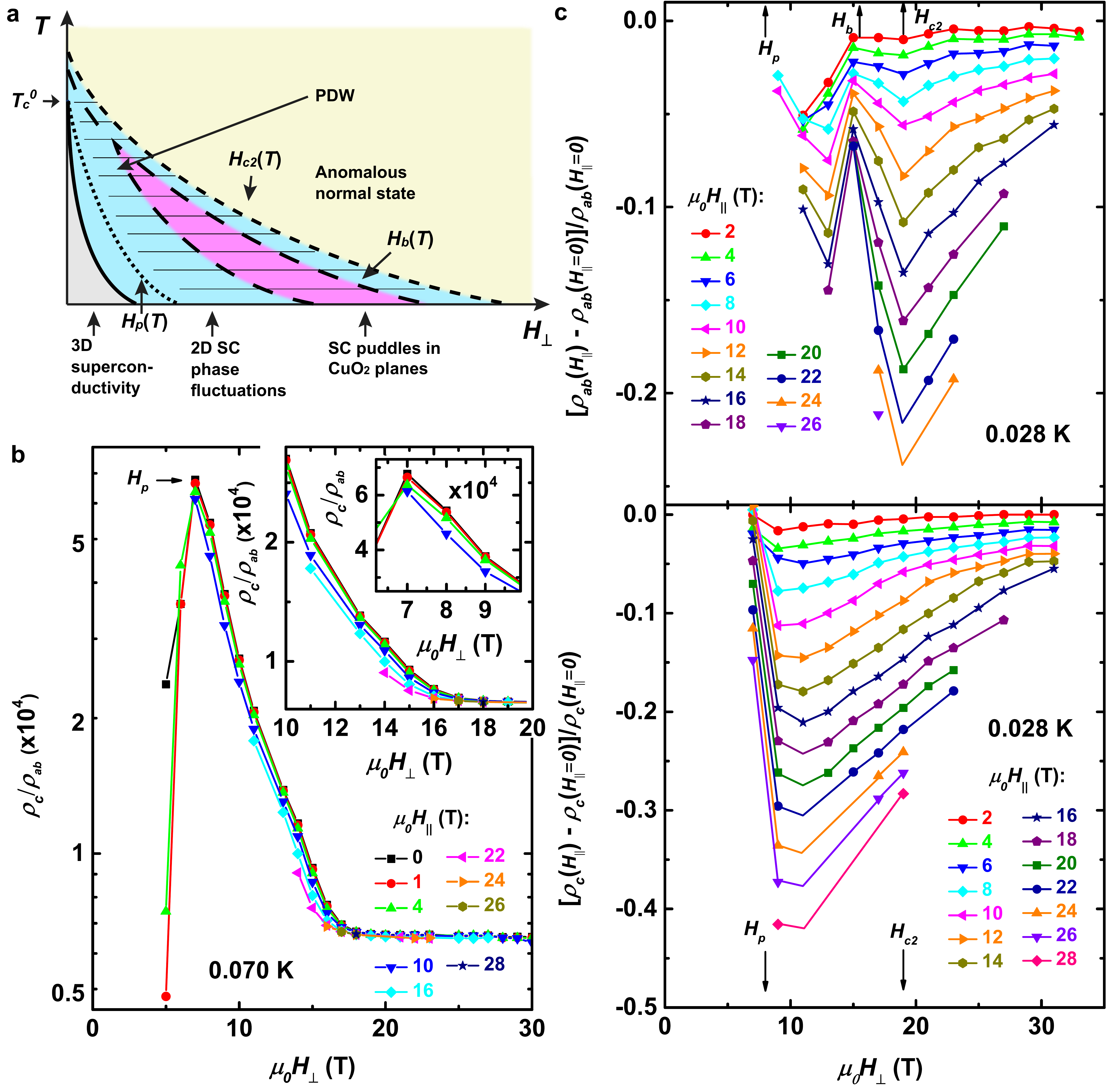}}
\caption{\textbf{Evidence for a PDW from anisotropic transport.}  \textbf{a}, 
Schematic $T$--$H_\perp$ phase diagram.  $H_\perp$ suppresses the 3D superconductivity (gray) and decouples (dotted line) the CuO$_2$ layers at $H_\perp=H_p(T)$.  Strong SC phase fluctuations persist in the planes up to $H_{c2}(T)$ (short-dashed line).  The behavior in the pink region, the precursors of which appear already in $H=0$ at $T>T_{c}^{0}$ (see dashed lines), is consistent with the presence of SC puddles in CuO$_2$ planes.  An additional, in-plane field enhances the interlayer coupling for $H_p(T) < H_\perp<H_{c2}(T)$, consistent with the presence of PDW correlations (thin hatched lines).  Except for the thick solid line, other lines do not represent phase boundaries, but correspond to finite-temperature crossovers.  \textbf{b}, $\rho_c/\rho_{ab}$ (for \lesco\, in-plane sample B1) vs $H_{\perp}$  for different $H_{\parallel}$, as shown, at $T=0.070$~K.  Larger inset: Enlarged view of the same data shows the suppression of the anisotropy by $H_{\parallel}$ for $H_p<H_{\perp}<H_{c2}$.  Smaller inset: $\rho_c/\rho_{ab}$ is reduced by $\sim$10\% near $H_p$ by $H_\parallel$ up to 10~T.  \textbf{c}, The corresponding $[\rho_{ab}(H_{\parallel})/\rho_{ab}(H_{\parallel}=0)-1]$ (top, sample B1) and $[\rho_{c}(H_{\parallel})/\rho_{c}(H_{\parallel}=0)-1]$ (bottom) vs $H_{\perp}$ at $T=0.028$~K for different $H_{\parallel}$, as shown.  In all panels, solid lines guide the eye.
}
\label{anisotropy}
\end{figure}

This picture is supported by the comparison of $\rho_c/\rho_{ab}$, as a function of $T$ and $H_\perp$, with the behavior of $\rho_{ab}(T)$ for a fixed $H_\perp$, as shown in Fig.~\ref{new-anis2} for both \lesco\, and \lnsco.  The $\rho_{ab}(T)$ data were extracted from the in-plane magnetoresistance (MR) measurements 
%
%
(ref.~\onlinecite{ZShi2019-Hc2}, Supplementary Fig.~2a; unless stated otherwise, the results are shown for \lesco\, sample ``B'', see Methods); the raw $\rho_c(H)$ data are shown in Supplementary Figs.~2b and 2c.  In Figs.~\ref{new-anis2}a and \ref{new-anis2}b, we also include $T_c(H_\perp)$, as well as $H_{peak}$, the position of the peak in the in-plane MR (see, e.g., Supplementary Fig.~2a), which corresponds\cite{ZShi2019-Hc2} to the upper critical field $H_{c2}$ in these materials (see also Supplementary Information).  Indeed, at a fixed $T$, $\rho_c/\rho_{ab}$ starts to increase as $H_\perp$ is reduced below $H_{peak}$.  This is followed by an enhancement of $\rho_c/\rho_{ab}$ near $H_\perp=H_{b}$, corresponding to the initial, metalliclike drop of $\rho_{ab}(T)$ as the SC transition is approached from the normal state for a fixed $H_\perp$ (Figs.~\ref{new-anis2}c and \ref{new-anis2}d).  The behavior of both materials is similar, except that the layer decoupling field $H_p(T)\gtrsim H_c(T)$ [or $T_c(H_\perp)$] in \lnsco, as expected\cite{Fradkin2015} for a stronger stripe order and frustration of interlayer coupling for $x\approx 1/8$.  Therefore, practically all the data in Figs.~\ref{new-anis2}c and \ref{new-anis2}d, i.e. for $H_\perp >H_p$, involve ``purely'' 2D physics, with no communication between the planes.  The striking splitting of the $\rho_{ab}(T)$ curves in both materials (ref.~\onlinecite{ZShi2019-Hc2}, Figs.~\ref{new-anis2}c and \ref{new-anis2}d), into either metalliclike (i.e. SClike) or insulatinglike, when the normal state sheet resistance $R_{\square/{\mathrm{layer}}}\approx R_{Q}$, where $R_{Q}=h/(2e)^{2}$ is the quantum resistance for Cooper pairs, further supports this conclusion: it agrees with the expectations for a 2D superconductor-insulator transition (SIT) driven by quantum fluctuations of the SC phase\cite{Fisher1990}.  In addition, as previously noted\cite{ZShi2019-Hc2}, the two-step $\rho_{ab}(T)$ is reminiscent of that in granular films of conventional superconductors and systems with nanoscale phase separation, including engineered Josephson junction arrays, where they are generally attributed to the onset of local (e.g. in islands or puddles) and global, 2D superconductivity.  Similarities to the behavior of various SC 2D systems\cite{CIQPT,Chen2018} thus suggest the formation of SC ``islands'' as $H_\perp$ is reduced below $H_b$ at a fixed $T$ (e.g. Figs.~\ref{new-anis2}a and \ref{new-anis2}b), i.e. at the initial, metalliclike drop of $\rho_{ab}(T)$ for a fixed $H_\perp$ ($H_b$ dashed line in Figs.~\ref{new-anis2}c and \ref{new-anis2}d).  Additional evidence in support of this interpretation, such as the $V$--$I$ that is characteristic of a viscous vortex liquid in the ``puddle'' regime, is discussed in Supplementary Information (also, Supplementary Figs.~3-5).  Therefore, at low $T$, the increasing $H_\perp$ destroys the superconductivity in the planes by quantum phase fluctuations of Josephson-coupled SC puddles.  The evolution of this ``puddle'' region with $T$ can be traced to the initial, metalliclike drop of $\rho_{ab}(T)$ at $T>T_{c}^{0}$ in $H=0$ (see $H_b$ dashed line in Figs.~\ref{new-anis2}c and \ref{new-anis2}d, and Supplementary Figs.~3 and 4).  Further increase of $H_\perp$ at low $T$ then leads to the loss of SC phase coherence in individual puddles and, eventually, transition to the high-field normal state.  These results are summarized in the sketch of the phase diagram, shown in Fig.~\ref{anisotropy}a.  

%
%

Our experiments are thus consistent with the presence of local PDW correlations (in ``puddles'') at $T>T_{c}^{0}$ in $H=0$, which are overtaken by the uniform $d$-wave superconductivity at low $T<T_{c}^{0}$.  In transport, the PDW SC order becomes apparent when the uniform $d$-wave order is sufficiently weakened by $H_\perp$: it appears \textit{beyond} the melting field of the vortex solid, within the vortex liquid regime, i.e. in the regime of strong 2D phase fluctuations.  Higher fields $H_p$ are needed to decouple the layers in \lesco\, than in \lnsco, since it is farther away from $x=1/8$.  In the $T\rightarrow 0$ limit and for even higher $H_\perp$ ($<H_{c2}$), the system seems to break up into SC puddles with the PDW order.   However, the final and key test of the presence of a PDW requires the application of a suitable perturbation, in particular $H_\parallel$, to reduce the interlayer frustration and decrease the anisotropy\cite{Fradkin2015}.

Therefore, we have performed angle-dependent measurements of both $\rho_{ab}(H)$ and $\rho_c(H)$, where the angle $\theta$ is between $H$ and the crystalline $c$ axis.  This has allowed us to explore the effect of in-plane fields $H_{\parallel}=H\sin\theta$ at different $H_{\perp}=H\cos\theta$, i.e. fields parallel to the $c$ axis, discussed above.  The angle-dependent $\rho_{ab}(H)$ was measured also on another \lesco\, sample (sample ``B1'', Methods; Supplementary Fig.~8); the results are qualitatively the same on both samples.  Figure~\ref{anisotropy}b illustrates the effect of $H_{\parallel}$ on $\rho_c/\rho_{ab}$ at low $T=0.070$~K on sample B1 (see Supplementary Figs.~9$\,$a-d for the raw $\rho_c$ and $\rho_{ab}$ data at different $T$).  Clearly, there is no effect of $H_{\parallel}$ for $H_{\perp}>H_{c2}(T=0.070$~K)$\approx17.5$~T.  Since $H_{\parallel}$ should break up Cooper pairs through the Zeeman effect, this confirms the absence of any observable remnants of superconductivity above the previously identified\cite{ZShi2019-Hc2} $H_{c2}$ ($\parallel c$).  In contrast, for $H_p\leqslant H_{\perp}<H_{c2}$, $H_{\parallel}$ \textit{reduces} the anisotropy, which is precisely what is expected in the presence of a PDW SC state if the dominant effect of $H_{\parallel}$ is to reorient the spin stripes\cite{Berg2007}.  

To understand exactly how $H_{\parallel}$ affects the anisotropy, we also investigate $\Delta\rho_{ab}=\rho_{ab}(H_{\parallel})-\rho_{ab}(H_{\parallel}=0)$ and $\Delta\rho_c=\rho_c(H_{\parallel})-\rho_c(H_{\parallel}=0)$ at different $H_{\perp}$ (Fig.~\ref{anisotropy}c and Supplementary Fig.~8d for sample B1; Supplementary Figs.~9$\,$e-h for sample B).  It is obvious that $\rho_{ab}$ is reduced by $H_{\parallel}$ for all $H_{\perp}$, which is the opposite of what would be expected if pair-breaking was dominant.  The suppression of $\rho_{ab}$ is weaker for those $H_{\perp}$ where the superconductivity is stronger, e.g. near $H_b\sim 15$~T in Fig.~\ref{anisotropy}c, and conversely, it is most pronounced above $H_{c2}$, indicating that the dominant effect of $H_{\parallel}$ is not related to superconductivity.  In fact, it occurs most strongly in the two regimes where $\rho_{ab}(H_{\perp})$ exhibits hysteretic behavior at low $T$ (Supplementary Figs.~3 and 6); the latter is attributed to the presence of domains with spin stripes (see also Supplementary Information and Supplementary Fig.~7).  This observation, therefore, further supports the conclusion that the main effect of $H_{\parallel}$ is the reorientation of spin stripes in every other plane\cite{Huecker2004,Huecker2008,Baek2015} (see also Supplementary Information).  The suppression of $\rho_{ab}$ by $H_{\parallel}$ seems to vanish at experimentally inaccessible $H_{\perp}$, where the anomalous, insulatinglike $\ln(1/T)$ dependence observed in the field-induced normal state also appears to vanish\cite{ZShi2019-Hc2}, suggesting that the origin of the $\ln(1/T)$ behavior might be related to the presence of short-range spin stripes.  As the spin stripes in every other plane are rotated by $H_{\parallel}$, in the PDW picture the interlayer frustration should be suppressed, leading to a decrease in $\rho_c$.  This is precisely what is observed (Fig.~\ref{anisotropy}c).  
The anisotropy ratio $\rho_c/\rho_{ab}$ is reduced (Fig.~\ref{anisotropy}b) because the effect of $H_{\parallel}$ on $\rho_c$ is relatively stronger than on $\rho_{ab}$.  Similar results are obtained in \lnsco\, (Supplementary Fig.~10): here the reduction in $\rho_c$ is weaker than in \lesco\, and $\rho_{ab}$ is not affected within the experimental resolution, both consistent with the stronger pinning of stripe order at $x=1/8$ (see also Supplementary Information).
Nevertheless, the reduction of $\rho_c/\rho_{ab}$ by $H_\parallel$ is comparable to that in \lesco\, (Fig.~\ref{anisotropy}b).  Therefore, by applying an      in-plane magnetic field, as proposed theoretically\cite{Berg2007,Fradkin2015}, our measurements confirm the presence of a PDW in both \lesco\, and \lnsco.  The effects of $H_\parallel$ are observable up to $T>T_{c}^{0}$ (i.e. $T\sim T_{SO}$ in \lesco: Supplementary Fig.~9), providing additional evidence for the PDW correlations in $H=0$ at $T>T_{c}^{0}$, as sketched in Fig.~\ref{anisotropy}a.

Finally, our results provide an explanation for the surprising, and \textit{a priori} counterintuitive, observation\cite{ZShi2019-Hc2} that $H_{c2}$ in \lnsco\, ($H_{c2}\sim 25$~T) is higher than in \lesco\, ($H_{c2}\sim 20$~T), even though its zero-field $T_{c}^{0}$ is lower because of stronger stripe correlations.  It is clear, though, that it is precisely because of the stronger stripe order and the presence of a more robust PDW SC state at $x\approx 1/8$ that the superconductivity persists to higher fields as $T\rightarrow 0$.

In summary, by probing the previously inaccessible high $H_\perp/T_{c}^{0}$ and $T\rightarrow 0$ regime dominated by quantum phase fluctuations and by testing a theoretical prediction, we have obtained evidence consistent with the existence of a PDW state in the La-214 family of cuprates with stripes.  Our observation of several signatures of a PDW in the regime with many vortices (i.e. a vortex liquid) is also consistent with the STM evidence\cite{Edkins2018} for a PDW order that emerges in vortex halos.  Since the observed PDW correlations extend only up to $T\ll T_{pseudogap}$ and not beyond $H_{c2}(T)$, our results do \textit{not} support a scenario in which the PDW correlations are responsible for the pseudogap.

\begin{methods}

\noindent\textbf{Samples.}  Several single crystal samples of La$_{1.8-x}$Eu$_{0.2}$Sr$_x$CuO$_4$ with a nominal $x=0.10$ and La$_{1.6-x}$Nd$_{0.4}$Sr$_x$CuO$_4$ with a nominal $x=0.12$ were grown by the traveling-solvent floating-zone technique\cite{Takeshita2004}.  The high homogeneity of the crystals was confirmed by several techniques, as discussed in detail elsewhere\cite{ZShi2019-Hc2}.  It was established that the samples were at least as homogeneous as those previously reported in the literature and, in fact, the disorder in our \lesco\, crystals was significantly lower than in other studies.  We note that the trivial possibility that the two-step SC transition observed at $H=0$ (e.g. Figs.~\ref{new-anis2}c and \ref{new-anis2}d for \lesco\, and \lnsco, respectively) may be due to an extrinsic inhomogeneity, e.g. the presence of two regions with different values of $T_{c}^{0}$, is clearly ruled out also by the behavior of $d\rho_{ab}/dT$ with $H_\perp$ (Supplementary Figs.~3a, 4, 8b).  In particular, both materials exhibit a reentrant metalliclike behavior at high $H_\perp$, below $H_{c2}$ (e.g. see the reentrant darker blue color band for \lnsco). This is the opposite of what is expected in case of two different $T_{c}^{0}$ values corresponding to different doping levels, where one would expect a gradual suppression of superconductivity with $H_\perp$, i.e. no reentrance.\\
\vspace*{-12pt}

\noindent The samples were shaped as rectangular bars suitable for direct measurements of the in-plane and out-of-plane resistance.  In \lesco, 
detailed measurements of $\rho_{ab}$ were performed on sample ``B'' with dimensions $3.06\times 0.53\times 0.37$~mm$^3$ ($a\times b\times c$); $\rho_c$ was measured on a bar with $0.34\times 0.41\times 1.67$~mm$^3$.  The in-plane \lnsco\, crystal with dimensions $3.82\times 1.19\times 0.49$~mm$^3$ was cut along the crystallographic [110] and [1$\bar{1}$0] axes, i.e. at a $45^{\circ}$ angle with respect to $a$ and $b$.  A bar with $0.21\times 0.49\times 3.9$~mm$^3$ ($a\times b\times c$) was used to measure $\rho_c$ in \lnsco.  The behavior of these samples remained astonishingly stable with time, without which it would have not been possible to conduct such an extensive and systematic study that required matching data obtained using different cryostats and magnets (see below) over the period of 2-3 years during which most of this study was performed, thus further attesting to the high quality of the crystals.  After $\sim 3$ years, the low-$T$ properties of sample B changed, resulting in a quantitatively different $T$--$H_\perp$ phase diagram (Supplementary Fig.~8b); this is why we consider it a different sample (``B1'').  The phase diagram of sample B1 seems to be intermediate to those of sample B (Supplementary Fig.~3a) and \lnsco\, (Supplementary Fig.~4).  Gold contacts were evaporated on polished crystal surfaces, and annealed in air at 700~$^{\circ}$C. The current contacts were made by covering the whole area of the two opposing sides with gold to ensure uniform current flow, and the voltage contacts were made narrow  to minimize the uncertainty in the absolute values of the resistance.  Multiple voltage contacts on opposite sides of the crystals were prepared.  The distance between the voltage contacts for which the data are shown is 1.53~mm for \lesco\, and 2.00~mm for \lnsco\, in-plane samples; 0.47~mm for \lesco\, and 1.26~mm for \lnsco\, out-of-plane samples.  Gold leads ($\approx 25~\mu$m thick) were attached to the samples using the Dupont 6838 silver paste, followed by the heat treatment at 450\,$^{\circ}$C in the flow of oxygen for 15 minutes. The resulting contact resistances were less than 0.1~$\Omega$ for \lesco\, (0.5~$\Omega$ for \lnsco) at room temperature.  The values of $T_{c}^{0}$ and the behavior of the samples did not depend on which voltage contacts were used in the measurements, reflecting the absence of extrinsic (i.e. compositional) inhomogeneity in these crystals.\\
\vspace*{-12pt}

\noindent $T_{c}^{0}$ was defined as the temperature at which the linear resistivity becomes zero.  For the in-plane samples, $T_{c}^{0}=(5.7\pm 0.3)$~K for \lesco\, and $T_{c}^{0}=(3.6\pm 0.4)$~K for \lnsco; the out-of-plane resistivity $\rho_c$ vanishes at $(5.5\pm 0.3)$~K for \lesco\, and $(3.4\pm 0.5)$~K for \lnsco.  In \lesco, $T_{SO}\sim 15$~K, $T_{CO}\sim 40$~K (ref.~\onlinecite{Fink2011}), and the pseudogap temperature $T_{pseudogap}\sim 175$~K (ref.~\onlinecite{Cyr2018}); in \lnsco, $T_{SO}\sim 50$~K, $T_{CO}\sim 70$~K (ref.~\onlinecite{Tranquada1996}), and $T_{pseudogap}\sim 150$~K (ref.~\onlinecite{Cyr2018}).  
\\
\vspace*{-12pt}

\noindent\textbf{Measurements.}   The standard four-probe ac method ($\sim 13$~Hz) was used for measurements of the sample resistance, with the excitation current (density) of 10~$\mu$A ($\sim 5 \times 10^{-3}$~A\,cm$^{-2}$ and $\sim 2 \times 10^{-3}$~A\,cm$^{-2}$ for \lesco\, and \lnsco, respectively) for the in-plane samples  and 10~nA ($\sim 7\times 10^{-6}$~A\,cm$^{-2}$ and $\lesssim 10^{-5}$~A\,cm$^{-2}$ for \lesco\, and \lnsco, respectively) for the out-of-plane samples.  $dV/dI$ measurements were performed by applying a dc current bias (density) down to 2~$\mu$A ($\sim 1 \times 10^{-3}$~A\,cm$^{-2}$ and $\sim 4 \times 10^{-4}$~A\,cm$^{-2}$ for \lesco\, and \lnsco\, in-plane samples, respectively) and a small ac current excitation $I_{ac}\approx 1~\mu$A ($\sim$ 13 Hz) through the sample and measuring the ac voltage across the sample.  For each value of $I_{dc}$, the ac voltage was monitored for 300~s and the average value recorded.  The relaxations of $dV/dI$ with time, similar to that in Supplementary Fig.~7, were observed only at the lowest $T\sim 0.016$~K.  Even then, the change of $dV/dI$ during the relaxation, reflected in the error bars for the $T=0.017$~K data in Supplementary Fig.~3c, was much smaller than the change of $dV/dI$ with $I_{dc}$.  The data that were affected by Joule heating at large dc bias were not considered.  To reduce the noise and heating by radiation in all measurements, a 1~k$\Omega$ resistor in series with a $\pi$ filter [5~dB (60~dB) noise reduction at 10~MHz (1~GHz)] was placed in each wire at the room temperature end of the cryostat.\\
\vspace*{-12pt}

\noindent The experiments were conducted in several different magnets at the National High Magnetic Field Laboratory: a dilution refrigerator (0.016 K $\leqslant$ T $\leqslant$ 0.7 K) and a $^{3}$He system (0.3 K $\leqslant$ T $\leqslant$ 35 K) in superconducting magnets ($H$ up to 18 T), using 0.1 -- 0.2~T/min sweep rates; a portable dilution refrigerator (0.02 K $\leqslant$ T $\leqslant$ 0.7 K) in a 35~T resistive magnet, using 1~T/min sweep rate; and a $^{3}$He system (0.3 K $\leqslant$ T $\leqslant$ 20 K) in a 31~T resistive magnet, using 1 -- 2~T/min sweep rates.  Below $\sim0.06$~K, it was not possible to achieve sufficient cooling of the electronic degrees of freedom to the bath temperature, a common difficulty with electrical measurements in the mK range.  This results in a slight weakening of the $\rho_{ab}(T)$ curves below $\sim0.06$~K for \textit{all} fields.  We note that this does not make any qualitative difference to the phase diagram (Supplementary Fig.~3a).  The fields were swept at constant temperatures, and the sweep rates were low enough to avoid eddy current heating of the samples.  The MR measurements with $H\parallel c$ were performed also by reversing the direction of $H$ to eliminate by summation any Hall effect contribution to the resistivity.  Moreover, since Hall effect had not been explored in these materials in large parts of the phase diagrams studied here, we have also carried out detailed measurements of the Hall effect; the results of that study will be presented elsewhere\cite{ZShi-Hall}.\\
\vspace*{-12pt}

\noindent The resistance per square per CuO$_2$ layer $R_{\square/\mathrm{layer}}=\rho_{ab}/l$, where $l=6.6$~\AA\, is the thickness of each layer.\\
\vspace*{-12pt}

\noindent\textbf{Data availability.}  The data that support the findings of this study are available from the corresponding author upon reasonable request.

\end{methods}

\bibliography{scibib}

\begin{thebibliography}{99}

\bibitem{Keimer2015} Keimer, B., Kivelson, S. A., Norman, M. R., Uchida, S. \& Zaanen, J.  From quantum matter to high-temperature superconductivity in copper oxides.  \textit{Nature} \textbf{518}, 179--186 (2015).

\bibitem{Comin2016} Comin, R. \& Damascelli, A. Resonant X-ray scattering studies of charge order in cuprates. \textit{Annu. Rev. Condens. Matter Phys.} {\bf 7}, 369--405 (2016).

\bibitem{Himeda2002} Himeda, A., Kato, T. \& Ogata, M. Stripe states with spatially oscillating $d$-wave superconductivity in the two-dimensional $t$-$t'$-$J$ model.  \textit{Phys. Rev. Lett.} \textbf{88}, 117001 (2002).

\bibitem{Berg2009-PRB} Berg, E., Fradkin, E. \& Kivelson, S. A.  Theory of the striped superconductor.  \textit{Phys. Rev. B} \textbf{79}, 064515 (2009).

\bibitem{Fradkin2015} Fradkin, E., Kivelson, S. A. \& Tranquada, J. M. Colloquium: Theory of intertwined orders in high temperature superconductors.  \textit{Rev. Mod. Phys.} {\bf 87}, 561--563 (2015).

\bibitem{Agterberg2019} Agterberg, D. F. \textit{et al.}  The physics of pair density waves: Cuprate superconductors and beyond.  \textit{Annu. Rev. Condens. Matter Phys.} \textbf{11}, 231--270 (2020).

\bibitem{Wen2012} Wen, J. \textit{et al.} Uniaxial linear resistivity of superconducting La$_{1.905}$Ba$_{0.095}$CuO$_4$ induced by an external magnetic field.  \textit{Phys. Rev. B} \textbf{85}, 134513 (2012).

\bibitem{Huecker2013} H\"ucker, M. \textit{et al.} Enhanced charge stripe order of superconducting La$_{2-x}$Ba$_x$CuO$_4$ in a magnetic field.  \textit{Phys. Rev. B} {\bf 87}, 014501 (2013).

\bibitem{Gerber2015} Gerber, S. \textit{et al.}  Three-dimensional charge density wave order in YBa$_2$Cu$_3$O$_{6.67}$ at high magnetic fields. \textit{Science} \textbf{350}, 949--952 (2015).

\bibitem{Berg2007} Berg, E. \textit{et al.}  Dynamical layer decoupling in a stripe-ordered high-$T_c$ superconductor.  \textit{Phys. Rev. Lett.} \textbf{99}, 127003 (2007).

\bibitem{Li2007} Li, Q., H\"ucker, M., Gu, G. D., Tsvelik, A. M. \& Tranquada, J. M. Two-dimensional superconducting fluctuations in stripe-ordered La$_{1.875}$Ba$_{0.125}$CuO$_4$.  \textit{Phys. Rev. Lett.} \textbf{99}, 067001 (2007).

\bibitem{Tajima2001} Tajima, S., Noda, T., Eisaki, H. \& Uchida, S.  $c$-axis optical response in the static stripe ordered phase of the cuprates.  \textit{Phys. Rev. Lett.} \textbf{86}, 500--503 (2001).

\bibitem{Stegen2013} Stegen, Z. \textit{et al.}  Evolution of superconducting correlations within magnetic-field-decoupled La$_{2-x}$Ba$_{x}$CuO$_4$ ($x=0.095$).  \textit{Phys. Rev. B} \textbf{87}, 064509 (2013).

\bibitem{SchafgansPRL2010} Schafgans, A. A. \textit{et al.}  Towards a two-dimensional superconducting state of La$_{2-x}$Sr$_x$CuO$_4$ in a moderate external magnetic field.  \textit{Phys. Rev. Lett.} \textbf{104}, 157002 (2010).

\bibitem{Lake2002} Lake, B., \textit{et al.} Antiferromagnetic order induced by an applied magnetic field in a high-temperature superconductor.  
\textit{Nature} \textbf{415}, 299--301 (2002).

\bibitem{Li2019} Li, Y. \textit{et al.}  Tuning from failed superconductor to failed insulator with magnetic field. \textit{Sci. Adv.} \textbf{5}, eaav7686 (2019).

\bibitem{Hamilton2018} Hamilton, D. R., Gu, G. D., Fradkin, E. \& Van Harlingen, D. J. Signatures of pair-density wave order in phase-sensitive measurements of La$_{2-x}$Ba$_x$CuO$_4$-Nb Josephson junctions and SQUIDs.  Preprint at https://arxiv.org/abs/1811.02048 (2018).

\bibitem{Edkins2018} Edkins, S. D. \textit{et al.}  Magnetic-field induced pair density wave state in the cuprate vortex halo.  \textit{Science} \textbf{364}, 976--980  (2019).

\bibitem{Huecker2004} H\"ucker, M.  \textit{et al.}  Dzyaloshinsky-Moriya spin canting in the low-temperature tetragonal phase of La$_{2-x-y}$Eu$_y$Sr$_x$CuO$_4$.  \textit{Phys. Rev. B} \textbf{70}, 214515 (2004).

\bibitem{Huecker2008} H\"ucker, M., Gu, G. D. \& Tranquada, J. M.  Spin susceptibility of underdoped cuprate superconductors: Insights from a stripe-ordered crystal.  \textit{Phys. Rev. B} \textbf{78}, 214507 (2008).

\bibitem{Baek2015}  Baek, S.-H. \textit{et al.}  Magnetic field induced anisotropy of $^{139}$La spin-lattice relaxation rates in stripe ordered La$_{1.875}$Ba$_{0.125}$CuO$_4$.  \textit{Phys. Rev. B} \textbf{92}, 155144 (2015).

\bibitem{Ando1995} Ando, Y., Boebinger, G. S., Passner, A., Kimura, T. \& Kishio, K. Logarithmic divergence of both in-plane and out-of-plane normal-state resistivities of superconducting La$_{2-x}$Sr$_x$CuO$_4$ in the zero-temperature limit.  \textit{Phys. Rev. Lett.} {\bf 75}, 4662--4665 (1995).

\bibitem{ZShi2019-Hc2} Shi, Z., Baity, P. G., Sasagawa, T. \& Popovi\'c, D.  Vortex phase diagram and the normal state of cuprates with charge and spin orders.   \textit{Sci. Adv.} \textbf{6}, eaay8946 (2020).

\bibitem{Fisher1990} Fisher, M. P. A. Quantum phase transitions in disordered two-dimensional superconductors.  \textit{Phys. Rev. Lett.} \textbf{65}, 923 (1990).

\bibitem{CIQPT} Dobrosavljevi\'c, V., Trivedi, N. \& Valles, J.M.  \textit{Conductor-Insulator Quantum Phase Transitions} (Oxford Univ. Press, 2012).

\bibitem{Chen2018} Chen, Z. \textit{et al.}  Carrier density and disorder tuned superconductor-metal transition in a two-dimensional electron system.  \textit{Nature Commun.} \textbf{9}, 4008 (2018).

\bibitem{Takeshita2004} Takeshita, N., Sasagawa, T., Sugioka, T., Tokura, Y., Takagi, H. Gigantic anisotropic uniaxial pressure effect on superconductivity within the CuO$_2$ plane of La$_{1.64}$Eu$_{0.2}$Sr$_{0.16}$CuO$_4$: Strain control of stripe criticality.  \textit{J. Phys. Soc. Jpn.} \textbf{73}, 1123--1126 (2004).
 
\bibitem{Fink2011} Fink, J. \textit{et al.}  Phase diagram of charge order in La$_{1.8-x}$Eu$_{0.2}$Sr$_x$CuO$_4$ from resonant soft x-ray diffraction.  \textit{Phys. Rev. B} {\bf 83}, 092503 (2011).

\bibitem{Cyr2018} O. Cyr-Choini\`ere, Pseudogap temperature $T^{\ast}$ of cuprate superconductors from the Nernst effect.  \textit{Phys. Rev. B} \textbf{97}, 064502 (2018).

\bibitem{Tranquada1996} Tranquada, J. M. \textit{et al.} Neutron-scattering study of stripe-phase order of holes and spins in \lnsco.  \textit{Phys. Rev. B} \textbf{54}, 7489--7499 (1996).

\bibitem{ZShi-Hall} Shi, Z., Baity, P. G., Sasagawa, T. \& Popovi\'c, D.  Magnetic field reveals zero Hall response in the normal state of stripe-ordered cuprates.  Preprint at https://arxiv.org/abs/1909.02491 (2019).

\end{thebibliography}

\bibliographystyle{Science}

\vspace*{6pt}

\noindent{\Large{\textbf{Acknowledgements}}}

\noindent We acknowledge helpful discussions with L. Benfatto, E. Berg, V. Dobrosavljevi\'c, E. Fradkin, S. A. Kivelson, J. M. Tranquada, K. Yang, and J. Zaanen.  This work was supported by NSF Grants Nos. DMR-1307075 and DMR-1707785, and the National High Magnetic Field Laboratory (NHMFL) through the NSF Cooperative Agreements Nos. DMR-1157490, DMR-1644779, and the State of Florida. \\
\vspace*{-6pt}

\noindent{\Large{\textbf{Author contributions}}}

\noindent Single crystals were grown and prepared by T.S.; Z.S., P.G.B., and J.T. performed the measurements and analysed the data; Z.S., P.G.B., J.T. and D.P. wrote the manuscript, with input from all authors; D.P. planned and supervised the investigation.\\
\vspace*{-6pt}

\noindent{\Large{\textbf{Additional information}}}

\noindent   
Supplementary information accompanies 
this paper.  Correspondence and requests for materials should be addressed to D.P.~(email: dragana@magnet.fsu.edu).\\
\vspace*{-6pt}

\noindent{\Large{\textbf{Competing financial interests}}}

\noindent The authors declare no competing interests.


\clearpage

\noindent\textbf{\Large{Supplementary Information}}
\vspace*{12pt}

\makeatletter
\makeatletter \renewcommand{\fnum@figure}{{\bf{\figurename~S\thefigure}}}
\makeatother

\setcounter{figure}{0}

\baselineskip=24pt

\noindent\textbf{Superconducting transition temperature and vortex phase diagram} 

$T_c(H_\perp)$ were determined as the temperatures at which the linear resistance $R\equiv\lim_{I_{dc}\rightarrow 0} V/I$ (or resistivity) becomes zero  ($V$ -- voltage, $I$ -- current).  In both \linebreak \lesco\, and \lnsco, $\rho_c$ and $\rho_{ab}$ vanish at the same temperature within the error, indicating the onset of 3D superconductivity.  In contrast, in striped La$_{1.875}$Ba$_{0.125}$CuO$_4$ in $H=0$, a 2D superconductivity was reported$^{11}$ to appear at a $T$ higher than the onset of 3D superconductivity, although it has been suggested$^4$ that higher precision measurements might reveal the same $T_{c}^{0}$ for both $\rho_c$ and $\rho_{ab}$. 

The position of the peak in $\rho_{ab}(H_\perp)$ (see, e.g., Fig.~S2a),  $H_\perp=H_{peak}(T)$, was found$^{23}$ to be of the order of the upper critical field ($H_{c2}$), i.e. the field scale corresponding to the closing of the SC gap.  Therefore, the superconductor with $T_c(H)>0$ (i.e. a $\rho_{ab}=0$ state) is separated from the normal state at $H_\perp>H_{peak}$ by a wide regime of SC phase fluctuations arising from the motion of vortices.  At low $T$, this regime exhibits non-Ohmic transport$^{23}$ consistent with the motion of vortices in the presence of disorder: it was thus identified$^{23}$ as a viscous vortex liquid with the zero freezing temperature, i.e. $T_c=0$.  
\vspace*{12pt}

\noindent\textbf{In-plane transport in perpendicular magnetic fields}

The in-plane $T$--$H_\perp$ phase diagrams of \lesco\, and \lnsco\, are shown in Figs.~S\ref{LESCO-phase}a and S\ref{LNSCO-phase}, respectively.  In order to display $\rho_{ab}(T)$ for \textit{all} $H_\perp$, we use the color maps.  The metalliclike, $d\rho_{ab}/dT>0$ regions where $\rho_{ab}(T>0)\neq 0$ (blue regions II  in Figs.~S\ref{LESCO-phase}a and S\ref{LNSCO-phase}), which exhibit non-Ohmic transport at low $T$, were identified$^{23}$ as a viscous vortex liquid with the zero freezing temperature, i.e. $T_c=0$. 

In \lesco, within the phase-fluctuations regime, there is clearly a region of pronounced insulatinglike ($d\rho_{ab}/dT<0$) behavior at low $T$ (region IV in Fig.~S\ref{LESCO-phase}a), with its precursors, i.e. the weakening of the metalliclike $T$ dependence, becoming visible already at $T\lesssim T_{c}^{0}$ (see also Fig.~S\ref{LNSCO-phase} for \lnsco), i.e. at $H_b(T)$ (see also Fig.~2).  The insulatinglike $d\rho_{ab}/dT$ that develops at low $T$, in region IV, is at least as strong as the one observed in the field-revealed normal state, i.e. for $H_\perp>H_{c2}\sim 20$~T (Fig.~S\ref{LESCO-phase}b).  By tracking the ``$h/4e^2$'' line where $R_{\square/{\mathrm{layer}}}$ changes from $R_{\square/{\mathrm{layer}}}<R_Q$ at lower $H_\perp$ and higher $T$, to $R_{\square/{\mathrm{layer}}}>R_Q$ at higher $H_\perp$ and lower $T$, we find that it has two branches (Figs.~S\ref{LESCO-phase}a and S\ref{LNSCO-phase}): while the upper one seems to form an upper limit for the presence of vortices$^{23}$, the lower one extrapolates roughly to the onset of region IV as $T\rightarrow 0$, suggesting that region IV may be related to the localization of Cooper pairs.  Indeed, although the range of $T$ and $H_\perp$ is limited, the scaling behavior of $\rho_{ab}(T,H_\perp)$  near the onset of region IV (Fig.~S\ref{scaling}) seems consistent with the presence of a $T=0$ SIT driven by quantum phase fluctuations in a disordered 2D system$^{23}$.  However, the $V$--$I$ measurements in region IV reveal a non-Ohmic \textit{increase} of $dV/dI$ with $I_{dc}$ (Fig.~S\ref{LESCO-phase}c), in contrast to the observations [S1] on the insulating side of the 2D SIT where $dV/dI$ decreases with $I_{dc}$.  On the other hand, a non-Ohmic increase of $dV/dI$ with $I_{dc}$ is consistent with the motion of vortices in the presence of disorder (i.e. a viscous vortex liquid) (refs.~23, [S2]).  The increase of $dV/dI$ with $I_{dc}$ is precisely the opposite of what would be expected in the case of simple Joule heating, confirming the presence of SC correlations, characteristic of a vortex liquid, in region IV.  Our results thus strongly suggest that region IV consists of SC puddles, with no inter-puddle phase coupling, in an insulatinglike, high-field normal-state background: at low $T$, the increasing $H_\perp$ destroys the superconductivity in the planes by quantum phase fluctuations of Josephson-coupled SC puddles.  The evolution of this region with $T$ can be traced to the initial, metalliclike drop of $\rho_{ab}(T)$ at $T>T_{c}^{0}$ in $H=0$ (see also $H_b$ dashed line in Figs.~2c and 2d).  In \lnsco, the lower branch of the ``$h/4e^2$'' line, together with $H_b(T)$, practically outlines the region of the weakened, metalliclike $\rho_{ab}(T)$.  Although the insulatinglike behavior is not observed, these results strongly suggest that it would ultimately emerge at even lower, experimentally inaccessible $T$, roughly in the $\sim10-20$~T field range, similar to \lesco.  

The weakening of the metalliclike $T$ dependence at intermediate $H_\perp$, which leads to the insulatinglike behavior in \lesco\, at low $T$ (region IV in Fig.~S\ref{LESCO-phase}), is manifested by the appearance of a ``shoulder'' in the in-plane MR curves for $H<H_{peak}$ (Figs.~S\ref{MR-hyst}a and S\ref{MR-LNSCO}a).  The shoulder in the MR becomes more noticeable with decreasing $T$ and, at very low $T\lesssim 0.05$~K, the MR in this range of fields becomes hysteretic (Fig.~S\ref{MR-hyst}a).  The size of the hysteresis grows with decreasing $T$ (Fig.~S\ref{MR-hyst}b), and the range of fields where it is observed, independent of the sweep rate, outlines the boundary of region IV (Fig.~S\ref{LESCO-phase}a) where $d\rho_{ab}/dT<0$.  In other words, the hysteretic MR is not observed at even higher $H_\perp$, where $d\rho_{ab}/dT>0$ (blue sliver in Fig.~S\ref{LESCO-phase}a).  Another hysteretic regime appears as the system enters the normal state (Fig.~S\ref{MR-hyst}a), but it is much less robust: its width in $H_\perp$ is reduced with decreasing sweep rate (Fig.~S\ref{MR-hyst}a inset; Fig.~S\ref{LESCO-phase}a shows the boundaries corresponding to 1~T/min).  In general, a hysteresis is a manifestation of the coexistence of phases, i.e. it indicates the presence of domains of different phases in the system.  Typical signatures of such systems include slow, nonexponential relaxations and memory effects, which are indeed observed here (Fig.~S\ref{relax}).  The hysteretic response to $H_\perp$, observed when the superconductivity is suppressed, is attributed to the presence of domains with spin stripes.
\vspace*{12pt}

\noindent\textbf{Effects of parallel magnetic fields on the spin structure}

In contrast to La$_{2-x}$Ba$_x$CuO$_4$, the magnetization of La$_{1.8-x}$Eu$_{0.2}$Sr$_x$CuO$_4$ and \linebreak La$_{1.6-x}$Nd$_{0.4}$Sr$_x$CuO$_4$
is dominated by the rare-earth ion (e.g. refs.~19, [S3]), so that studies of the Cu spin magnetism are difficult and scarce in these compounds.  Nevertheless, it is known that, in the low-temperature tetragonal phase of antiferromagnetic La$_{1.8}$Eu$_{0.2}$CuO$_4$, an in-plane magnetic field ($H\parallel b$) of $\sim 6$~T leads to a spin-flop transition$^{19}$ of the Cu spin moments in every other plane.  Moreover, this spin-flop transition field is roughly the same as that in La$_{1.875}$Ba$_{0.125}$CuO$_4$ for $T<T_{SO}$ (ref.~20).  Since the structure of these three materials is similar [S4], it is thus likely that the spin-flop transition occurs also in La$_{1.8-x}$Eu$_{0.2}$Sr$_x$CuO$_4$ and La$_{1.6-x}$Nd$_{0.4}$Sr$_x$CuO$_4$ near $x=1/8$ at comparable fields.  La$_{2-x}$Sr$_{x}$CuO$_4$ with $x=0.115$, for example, also exhibits a spin-flop transition at a similar field $\approx 7.5$~T [S5].
We note that, in case of $H\parallel [110]$, the transition is broader: spins in all planes continuously rotate until staggered moment is again perpendicular to the field, so there is no \textit{sharp} spin-flop transition$^{20,21}$.  In all cases, however, the reorientation of spins under the influence of $H\parallel ab$ should enhance Josephson coupling between layers, within the PDW picture.

In the $H\parallel [110]$ configuration, Josephson coupling between layers may be enhanced also by another mechanism, namely by the field partially compensating for the momentum mismatch between the layers; that mechanism would reduce $\rho_c$, but it would have no effect on $\rho_{ab}$  [S6].  In our in-plane \lnsco\, crystal, which was cut at a $45^{\circ}$ angle with respect to $a$ and $b$ axes (Methods), we do not see, indeed, any observable effect of $H\parallel [110]$ on $\rho_{ab}$ (Fig.~S\ref{LNSCO-angle}a).  This also implies that, in contrast to \lesco, (Fig.~3c, top), the effect of spin reorientation on $\rho_{ab}$ is too weak to be observed within the experimental resolution.  We note that, at the same time, in our out-of-plane \lnsco\, crystal for which $H\parallel [100]$ (Methods) and in which the mechanism of ref.~[S6] thus cannot play a role, the reduction in $\rho_c$ (Fig.~S\ref{LNSCO-angle}b) is also weaker than in \lesco.  The weaker spin reorientation effect of $H_{\parallel}$ on $\rho_c$ and $\rho_{ab}$ in \lnsco\, than in \lesco\, is, therefore, attributed to the stronger pinning of stripe order at $x=1/8$.  

%
\begin{figure}[!tb]
\centerline{\includegraphics[width=0.7\textwidth]{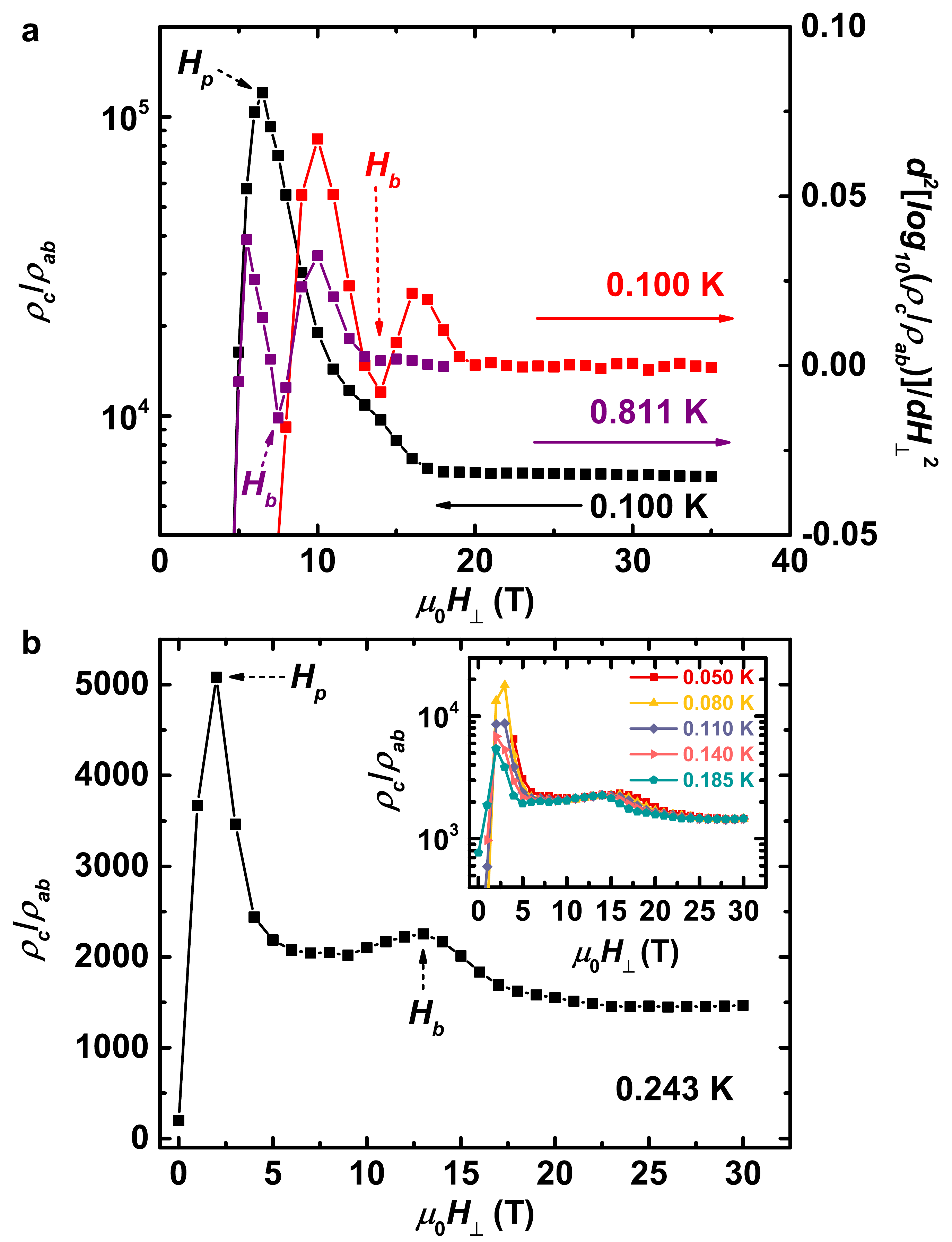}}
\caption{\textbf{Methods to determine characteristic fields in the $\bm{H_\perp}$ dependence of the anisotropy ratio $\bm{\rho_{c}/\rho_{ab}}$.}  
\textbf{a,} \lesco.  The anisotropy ratio $\rho_c/\rho_{ab}$ (black symbols, left axis) vs $H_\perp$ at $T=0.100$~K on a semi-log scale.  Red and purple symbols (right axis) show the second derivative $d^{2}[\log(\rho_c/\rho_{ab})]/dH_{\perp}^2$ for $T=0.100$~K and $T=0.811$~K, respectively.  Solid lines guide the eye.  $H_b$ is defined as the minimum in the second derivative, as shown.  Clearly, $H_b$ remains strongly pronounced even at a fairly high $T$.  The analysis was repeated for different $T$.  \textbf{b,} \lnsco; $\rho_c/\rho_{ab}$ at $T=0.243$~K.  Although the absolute value of the anisotropy is relatively low, as noted previously for La-214 cuprates with $x\approx 1/8$ [e.g. Berg \textit{et al.}, New J. Phys. \textbf{11}, 115004 (2009)], the enhancement of $\rho_c/\rho_{ab}$ at $H_b$ is clearly observed already in the raw data.  Inset:  $\rho_c/\rho_{ab}$ vs $H_\perp$ for several $T$.}
\label{method}
\end{figure}

\clearpage

%
\begin{figure}
\centerline{\includegraphics[width=0.65\textwidth]{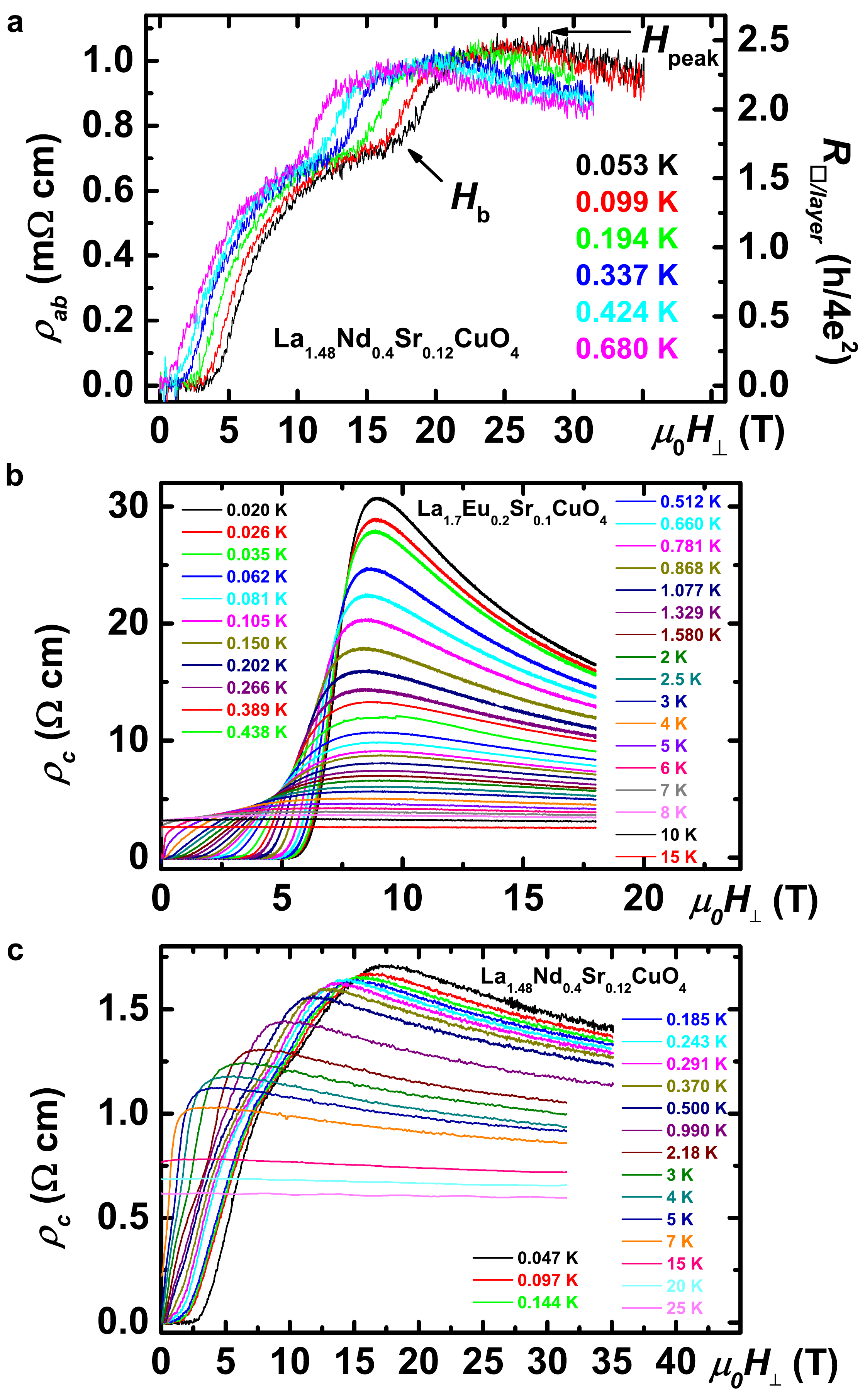}}
\caption{\textbf{The dependence of the in-plane and out-of-plane resistivity on $\bm{H\parallel c}$ (i.e. $\bm{H_\perp}$).}   \textbf{a}, $\rho_{ab}$ vs $H_\perp$ at several $T<T_{c}^{0}$ in  La$_{1.48}$Nd$_{0.4}$Sr$_{0.12}$CuO$_4$.  At low $T$, $\rho_{ab}(H_\perp)$ exhibits a peak at $H_\perp=H_{peak}(T)$.  $H_b(T)$ corresponds to the establishment of SC correlations in the planes, as the SC transition is approached from a high-field normal state; see also Figs. 1 and 2.  The right axis shows the corresponding  $R_{\square/\mathrm{layer}}$ in units of quantum resistance for Cooper pairs, $R_{Q}=h/(2e)^{2}$.  \textbf{b} and \textbf{c}, $\rho_{c}$ vs $H_\perp$ for several $T$ in \lesco\, and \lnsco, respectively.
}
\label{MR-LNSCO}
\end{figure}
%

\clearpage

%
\begin{figure}[!tb]
\centerline{\includegraphics[width=0.83\textwidth]{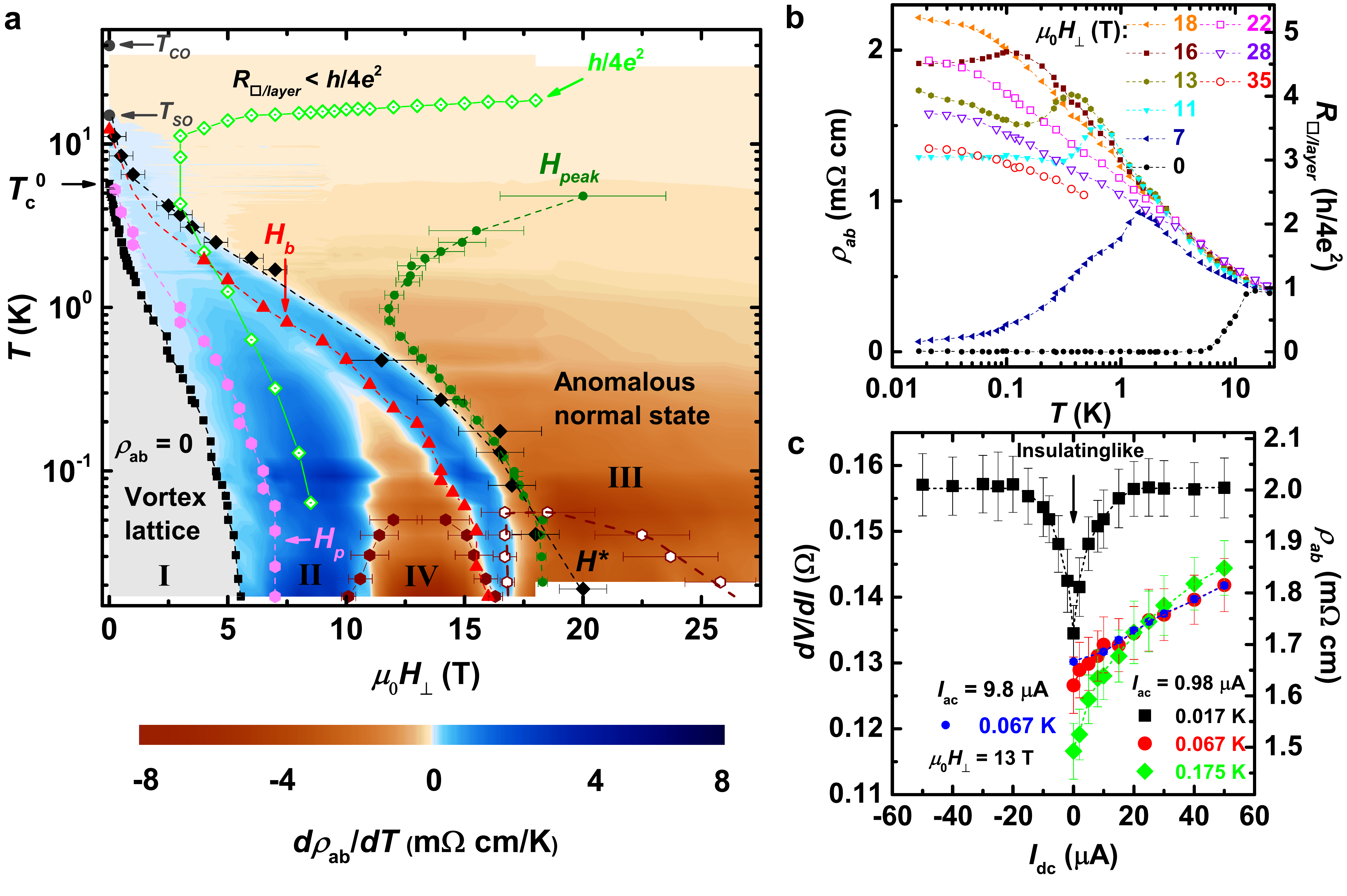}}
\caption{\textbf{$\bm{T}$--$\bm{H_\perp}$ phase diagram of La$\bm{_{1.7}}$Eu$\bm{_{0.2}}$Sr$\bm{_{0.1}}$CuO$\bm{_4}$.}  \textbf{a}, Black squares: $T_c(H_\perp)$; $\rho_{ab}=0$ for all $T<T_c(H_\perp)$ (region I).  The color map: $d\rho_{ab}/dT$; in the viscous vortex liquid (II), $T_{c}=0$.  The dark brown dots mark the  $(T,H_\perp)$ range in which the MR hysteresis, independent of the field sweep rate, is observed; the dark brown open dots show the boundary of the hysteretic regime observed with a 1~T/min sweep rate (Fig.~S\ref{MR-hyst}a); the error bars reflect the uncertainty in $\rho_{ab}$ due to $T$ fluctuations and the experimental resolution for estimating the onsets of bifurcation.  Green dots: $H_{peak}(T)\sim H_{c2}(T)$.   Black diamonds: $H^{\ast}(T)$, the boundary between non-Ohmic $V$--$I$ for $H_\perp<H^{\ast}$ and Ohmic behavior found at $H_\perp>H^{\ast}$.   Region III is the $H_\perp$-revealed normal state.  Open green diamonds: the $h/4e^2$ line.  Pink dots: $H_{p}(T)$; red triangles: $H_{b}(T)$.  All dashed lines guide the eye.  $T_{SO}(H=0)$ and $T_{CO}(H=0)$ are also shown; both spin and charge stripes are known to be enhanced by $H_\perp$.  \textbf{b}, $\rho_{ab}(T)$ for several $0\leq H_\perp\leq 35$~T;  dashed lines guide the eye.  $d\rho_{ab}/dT<0$ in region IV, e.g. for $H_\perp=13$~T and $T<0.1$~K, is comparable to that observed in the normal state ($H_\perp>20$~T), e.g. for $H_\perp=28$~T and $H_\perp=35$~T.  In region III, $\rho_{ab}\propto\ln(1/T)$ is obeyed$^{23}$ at least down to $\sim 0.06$--0.07~K, below which it was not possible to achieve sufficient cooling of the electronic degrees of freedom to the bath temperature, a common difficulty with electrical measurements in the mK range; this results in a slight weakening of the $\rho_{ab}(T)$ below $\sim0.06$~K for \textit{all} fields.  \textbf{c}, $dV/dI$ vs $I_{dc}$ for several $T$ at $H_\perp=13$~T (region IV); $I_{ac}\approx 1~\mu$A, but the data taken at $T=0.067$~K show that the same result is obtained, within the error, with $I_{ac}\approx 1~\mu$A and $I_{ac}\approx 10~\mu$A.  The temperature dependence of the linear resistance ($dV/dI$ for $I_{dc}\rightarrow 0$) is insulatinglike.  Dashed lines guide the eye.  
}
\label{LESCO-phase}
\end{figure}
%

\clearpage

%
\begin{figure}
\centerline{\includegraphics[width=0.83\textwidth]{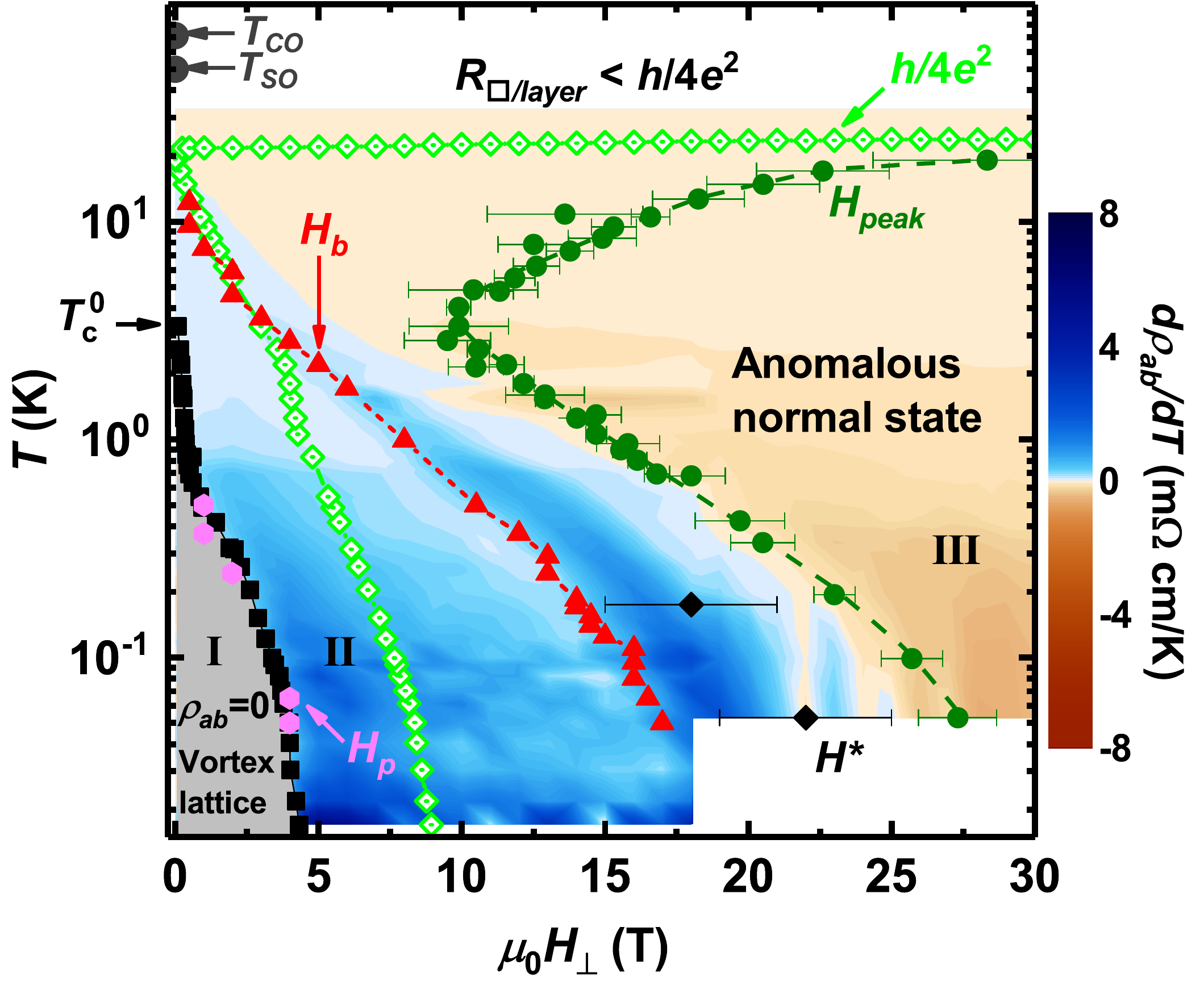}}
\caption{\textbf{In-plane transport $\bm{T}$--$\bm{H_\perp}$ phase diagram of La$\bm{_{1.48}}$Nd$\bm{_{0.4}}$Sr$\bm{_{0.12}}$CuO$\bm{_4}$.}  Black squares: $T_c(H)$; $\rho_{ab}=0$ for all $T<T_c(H)$ [region I; $T_c(H)>0$].  The color map: slopes $d\rho_{ab}/dT$, clearly indicating a weakening (lighter blue color) of the metalliclike behavior at intermediate fields, i.e. within the viscous vortex liquid, for which $T_{c}=0$ (region II).  $H_{peak}(T)\sim H_{c2}(T)$ (green dots) represent fields above which the MR changes from positive to negative.  Region III is the $H$-induced normal state.  Open green diamonds: the $h/4e^2$ line.  Pink dots: $H_{p}(T)$; red triangles: $H_{b}(T)$.  In contrast to \lesco, here the layer decoupling field $H_p(T)\gtrsim H_c(T)$ (black squares), consistent with a stronger stripe order for $x\approx 1/8$.  As in \lesco, the boundary of the weakened $\rho_{ab}(T)$ at intermediate fields is outlined by $H_b$ and, roughly, by the $h/4e^2$ line.  These results suggest that the insulatinglike regime would emerge at even lower, experimentally inaccessible $T$, in the $\sim 10-20$~T field range.  All dashed lines guide the eye.  Black diamonds: $H^{\ast}(T)$ represent the boundary$^{23}$ between non-Ohmic $V$--$I$ for $H_\perp<H^{\ast}$ and Ohmic behavior found at $H_\perp>H^{\ast}$.  $H=0$ values of $T_{SO}$ and $T_{CO}$ are also shown; both spin and charge stripes are known to be enhanced by $H_\perp$ (see main text).
}
\label{LNSCO-phase}
\end{figure}
%

\clearpage

%
\begin{figure}
\centerline{\includegraphics[width=\textwidth]{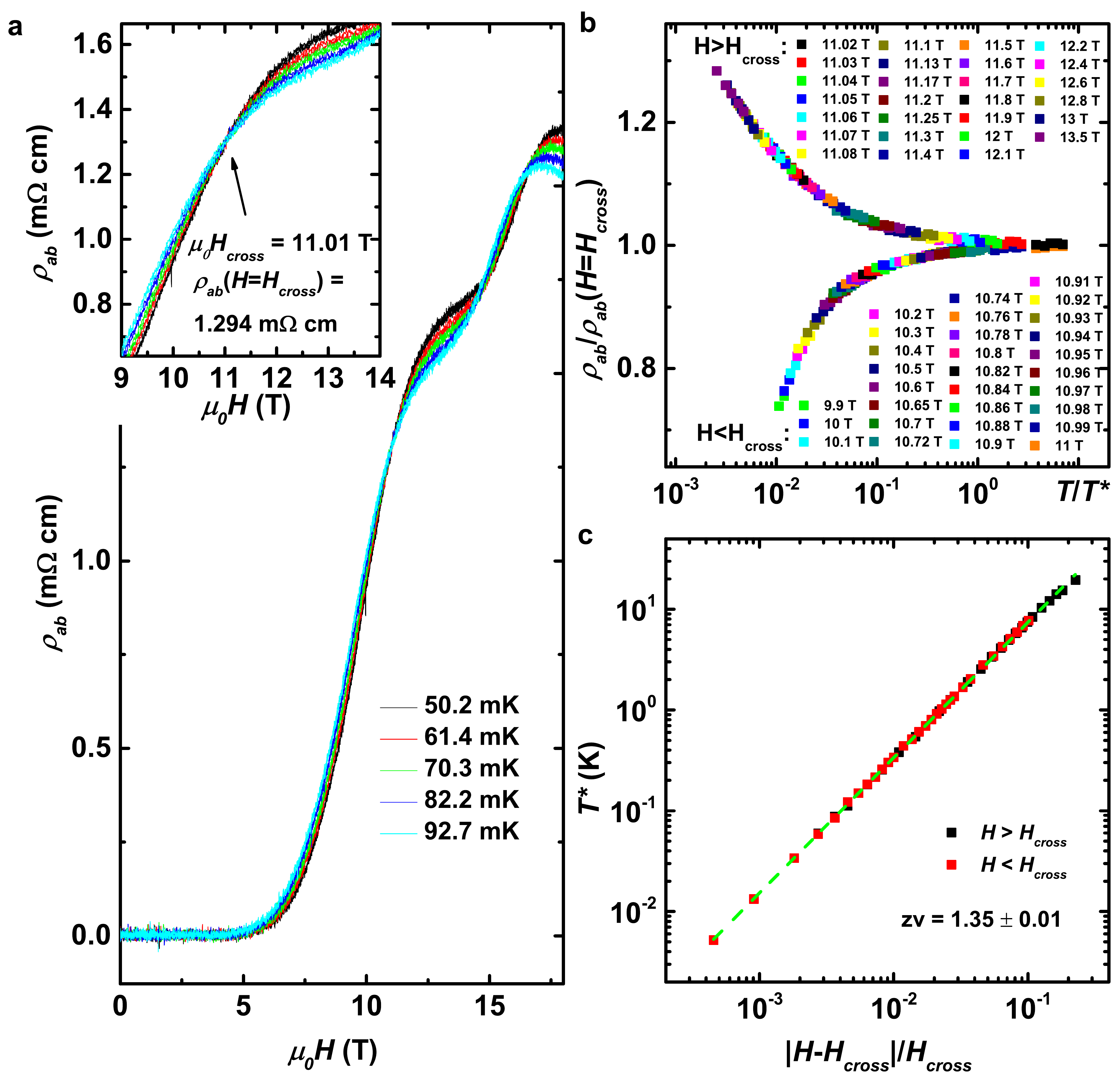}}
\caption{\textbf{Scaling of the in-plane resistivity $\bm{\rho_{ab}(T,H)}$ near the onset of region IV in Fig.~S\ref{LESCO-phase}a in La$\bm{_{1.7}}$Eu$\bm{_{0.2}}$Sr$\bm{_{0.1}}$CuO$\bm{_4}$; $\bm{H\equiv H_\perp}$.}  \textbf{a,} Isothermal $\rho_{ab}(H)$ curves at low $T$ show the existence of a $T$-independent crossing point (inset) at $\mu_{0}H_{cross}=11.01$~T and $\rho_{ab}(H=H_{cross})=1.294$~m$\Omega$cm (or $R_{\square/\mathrm{layer}}\approx 3~h/4e^2$).  \textbf{b,} Scaling of the data in \textbf{a} with respect to a single variable $T/T^{\ast}$; here, $\rho_{ab}(T,H)=\rho_{ab}(H=H_{cross})f(T/T^{\ast})$, i.e. the resistivity data for different $H$ can be collapsed onto a single function by rescaling the temperature.  \textbf{c,} The scaling parameter $T^{\ast}$ as a function of $|\delta|=|H-H_{cross}|/H_{cross}$ on both sides of $H_{cross}$.  The dashed line is a linear fit with the slope $z\nu=1.35\pm 0.01$, as shown; $T^{\ast}\propto |\delta|^{z\nu}$.
}
\label{scaling}
\end{figure}
%

\clearpage

%
\begin{figure}[!tb]
\centerline{\includegraphics[width=0.98\textwidth]{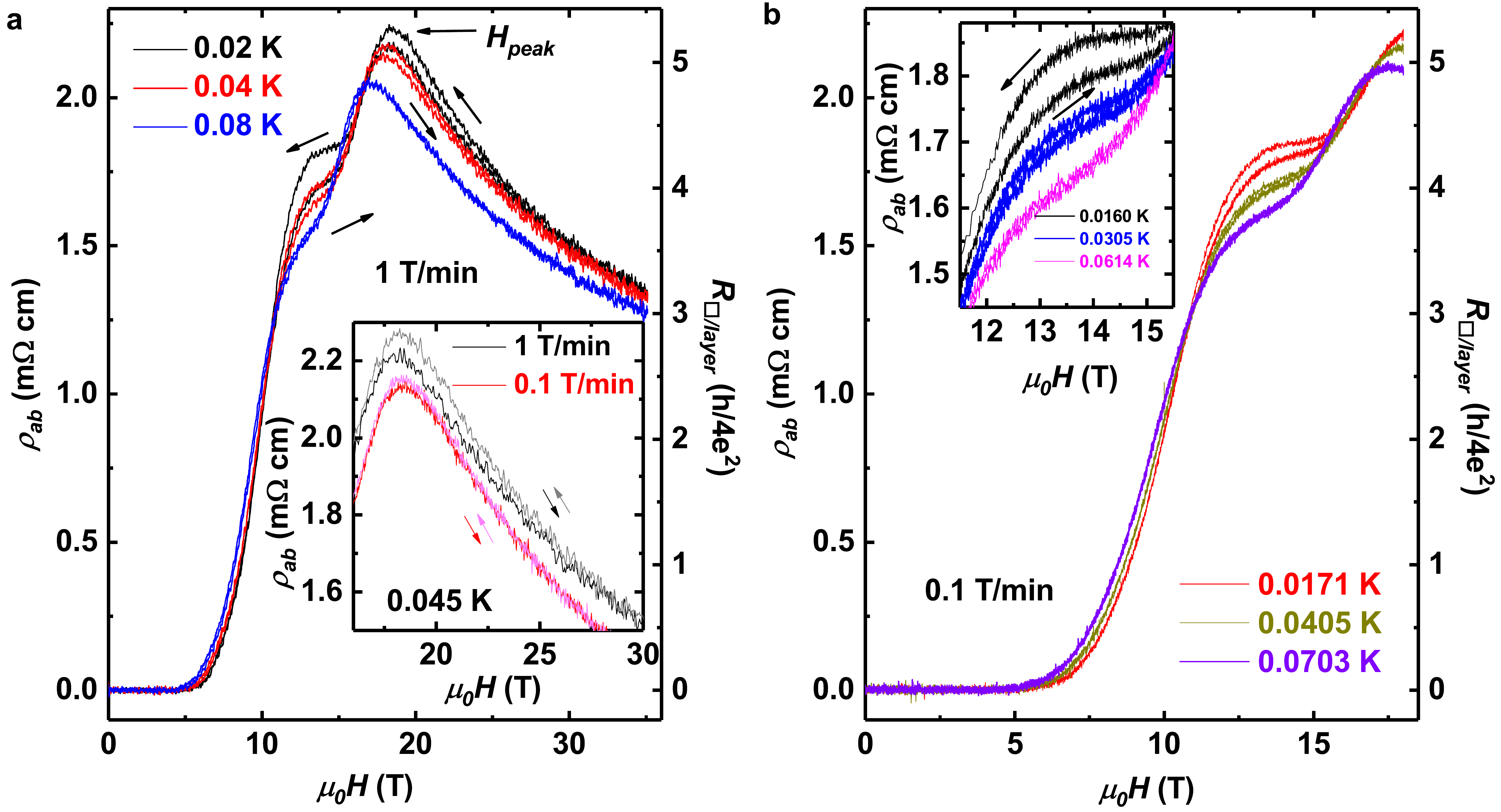}}
\caption{\textbf{Linear in-plane resistivity of La$\bm{_{1.7}}$Eu$\bm{_{0.2}}$Sr$\bm{_{0.1}}$CuO$\bm{_4}$ vs $\bm{H\parallel c}$.} 
\textbf{a},  At low $T$, $\rho_{ab}(H)$ exhibits a sharp peak at $H=H_{peak}(T)$ and two hysteretic regimes: one occurs near a shoulder below the peak (region IV in Fig.~S\ref{LESCO-phase}a) and the other starts at $H\sim H_{peak}$.  The width in $H$ of the lower-field hysteretic region is the same for sweep rates between 1~T/min, shown here, and 0.1~T/min (see \textbf{b}).  Inset: The higher-field hysteresis is less robust, as its width is reduced with decreasing sweep rate.  The 0.1~T/min trace, which shows a small hysteresis near $H_{peak}$, is shifted down by 0.12~m$\Omega$cm for clarity.  Arrows show the direction of field sweeps.  \textbf{b}, The hysteretic, insulatinglike region IV is surrounded by the regimes of metallic behavior.  Inset: The hysteresis is suppressed with increasing $T$. 
}
\label{MR-hyst}
\end{figure}
%

%
\begin{figure}
\centerline{\includegraphics[width=0.86\textwidth]{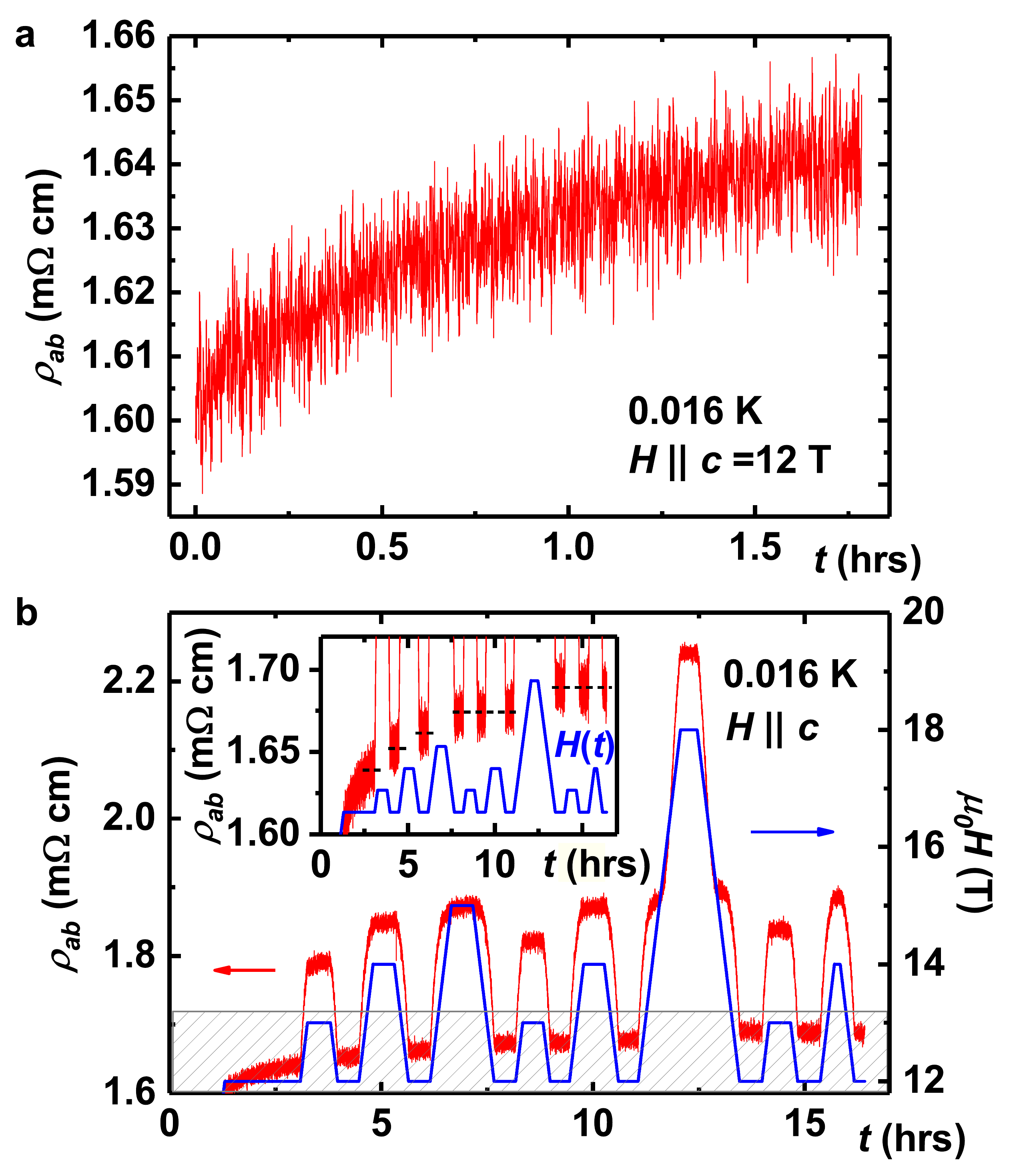}}
\caption{\textbf{Nonequilibrium dynamics in region IV of the La$\bm{_{1.7}}$Eu$\bm{_{0.2}}$Sr$\bm{_{0.1}}$CuO$\bm{_4}$ phase diagram in Fig.~S\ref{LESCO-phase}a.} \textbf{a,} ${\rho}_{ab}$ exhibits slow, nonexponential relaxations with time $t$: here it continues to relax for hours after the magnetic field reaches 12 T at $T=0.016$~K.  \textbf{b,} At a fixed $T=0.016$~K, ${\rho}_{ab}$ (red; left axis) is measured as a function of time as $H_\perp$ is changed between 12 T and different higher fields (blue; right axis).  This protocol allows a comparison of ${\rho}_{ab}$ values obtained at the same ${\mu}_{0}H_\perp=12$~T but with a different magnetic history. Inset: Enlarged shaded area of the main plot shows that ${\rho}_{ab}({\mu}_{0}H_\perp=12$~T) is determined by the highest $H_\perp$ applied previously: the system acquires a memory of its magnetic history. Dashed lines guide the eye.}
\label{relax}
\end{figure}

\clearpage

\begin{figure}
\centerline{\includegraphics[width=\textwidth]{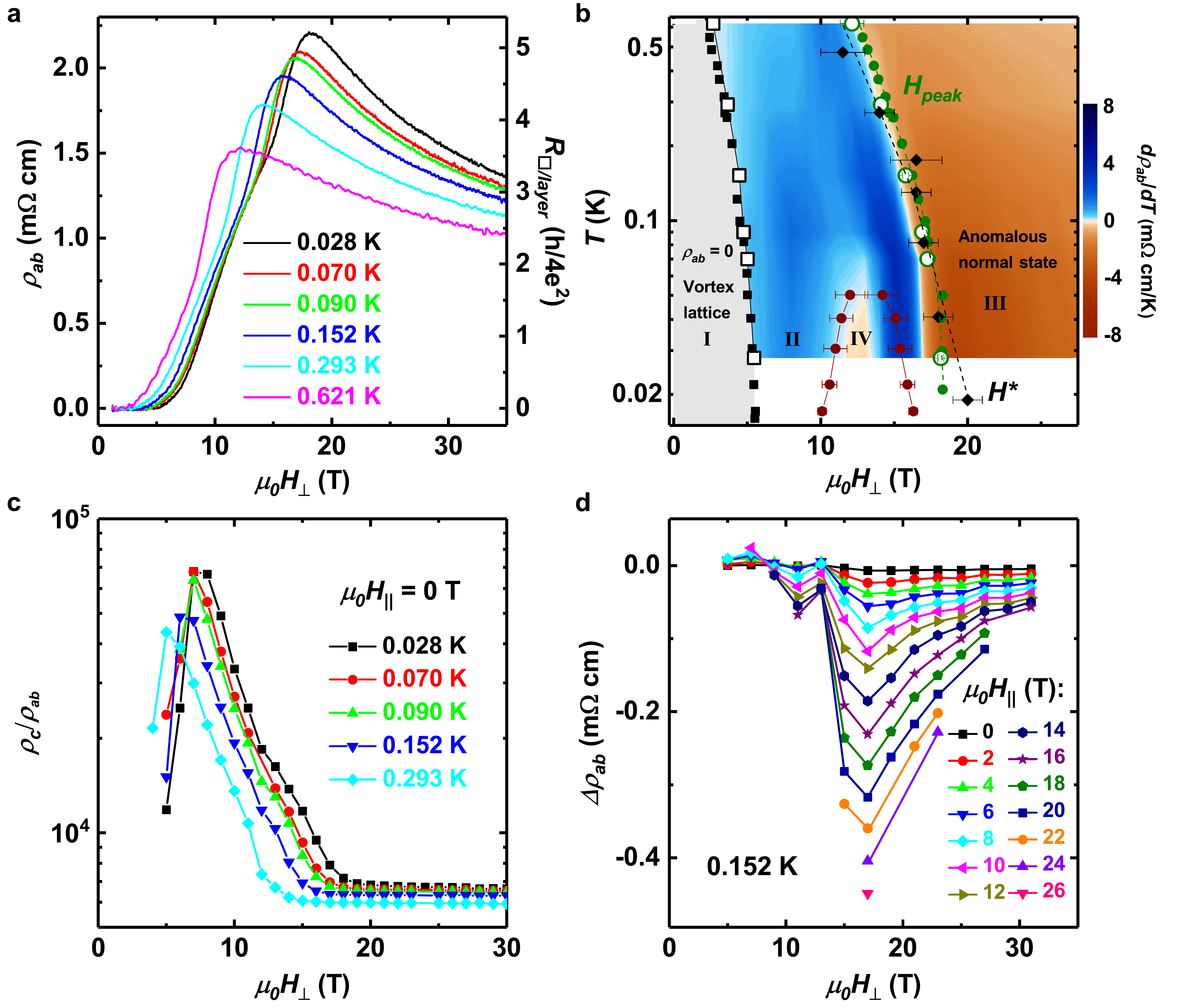}}
\caption{\textbf{In-plane La$\bm{_{1.7}}$Eu$\bm{_{0.2}}$Sr$\bm{_{0.1}}$CuO$\bm{_4}$ sample B1.}  \textbf{a,} $\rho_{ab}$ vs $H_{\perp}$ (i.e. $H\parallel c$) for several $T$, as shown.  \textbf{b,} In-plane transport $T$--$H$ phase diagram with $H\parallel c$ axis.  The color map shows $d\rho_{ab}/dT$ on the same scale as that in Fig.~S\ref{LESCO-phase}a for sample B.  Open black squares and open green dots represent $T_c(H)$ and $H_{peak}(T)$, respectively.  For comparison, solid symbols show the corresponding values for sample B; dark brown dots show the boundary of the hysteretic regime (region IV) in sample B.  While the values of $T_c(H)$ and $H_{peak}(T)$ in B and B1 match within error, the insulatinglike region IV is clearly suppressed to lower $T$ in sample B1 but, at the same time, the reentrant vortex liquid regime is more pronounced.  For completeness, solid diamonds show the values of $H^{\ast}(T)$, the boundary between non-Ohmic and Ohmic transport, for sample B.  \textbf{c,}  $\rho_c/\rho_{ab}$ vs $H\parallel c$ at different $T$, as shown.  Solid lines guide the eye.  \textbf{d,} The suppression of the in-plane resistivity by $H_{\parallel}$, $\Delta\rho_{ab}=\rho_{ab}(H_{\parallel})-\rho_{ab}(H_{\parallel}=0)$, for different $H_{\parallel}$, as shown, as a function of $H_{\perp}$ at $T=0.152$~K.  Solid lines guide the eye.
}
\label{B1}
\end{figure}
%

\clearpage

\begin{figure}
\centerline{\includegraphics[width=0.85\textwidth]{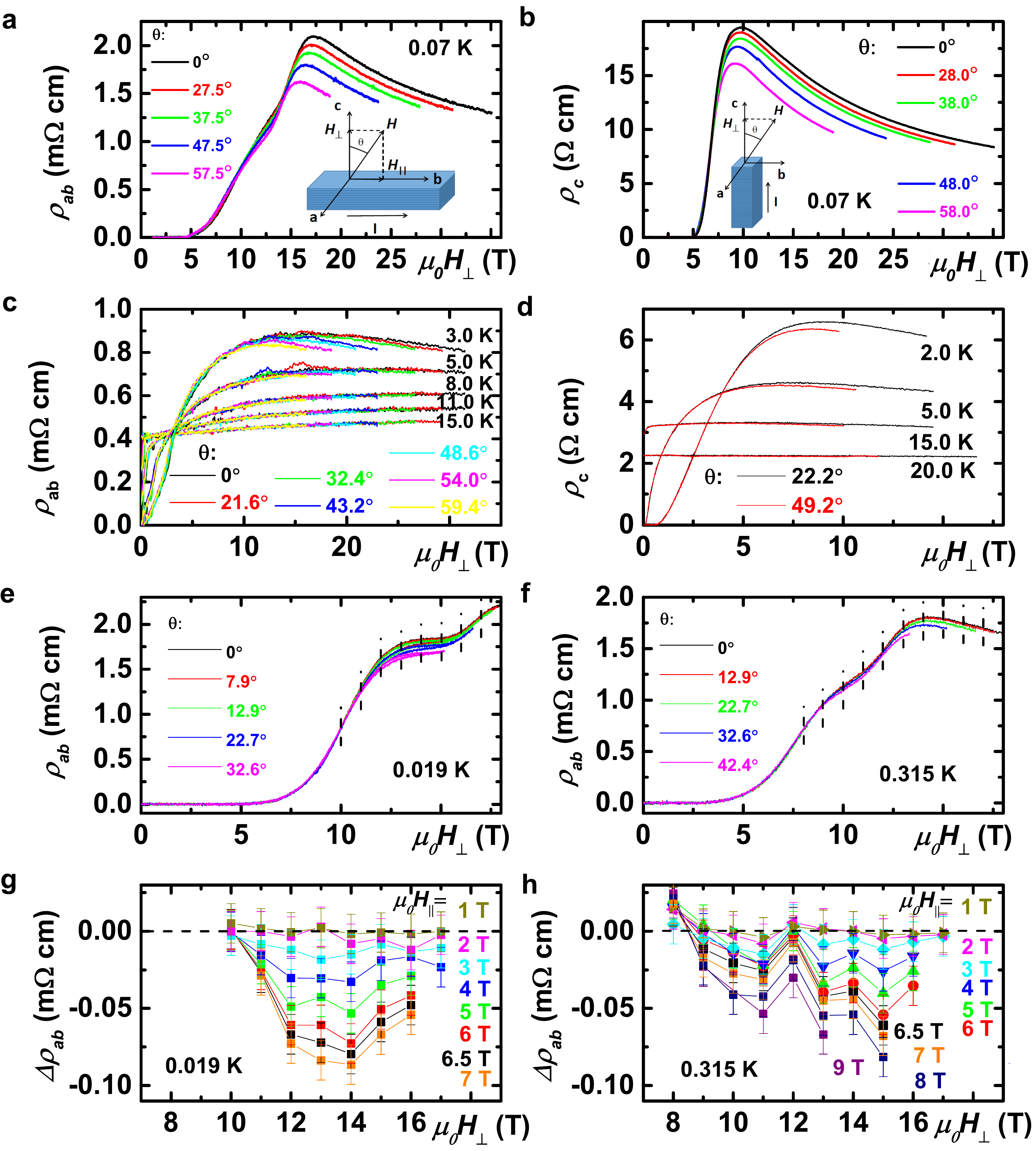}}
\caption{\textbf{Angle-dependent transport in La$\bm{_{1.7}}$Eu$\bm{_{0.2}}$Sr$\bm{_{0.1}}$CuO$\bm{_4}$ vs $\bm{H_{\perp}=H\parallel c}$ axis.}  \textbf{a,} $\rho_{ab}$ (sample B1), and \textbf{b,} $\rho_c$, at $T=0.07$~K and for different angles $\theta$, as shown; the uncertainty $\Delta\theta= 0.5^{\circ}$.  The sketches show the field orientation with respect to the sample axes and the current flow.  \textbf{c,} $\rho_{ab}$ (sample B), and \textbf{d,} $\rho_c$, at higher $T$ and for different angles $\theta$, as shown.  While the effect of $H_{\parallel}$ on $\rho_{ab}$ vanishes at $\sim 5$~K, i.e. at $T\sim T_{c}^{0}$ (\textbf{c}), in $\rho_c$ it vanishes at $\sim 15$~K, i.e. at $T\sim T_{SO}$ (\textbf{d}); in both cases, this seems to be related to the vanishing of the peak in the MR.  \textbf{e-h,} Sample B.  $\rho_{ab}(H_{\perp})$ for different angles $\theta$, as shown, at $T=0.019$~K (\textbf{e}) and $T=0.315$~K (\textbf{f}).  Vertical dashed lines indicate the values of $H_{\perp}$ used in figures \textbf{g} and \textbf{h}.  \textbf{g} and \textbf{h} show $\Delta\rho_{ab}=\rho_{ab}(H_{\parallel})-\rho_{ab}(H_{\parallel}=0)$, i.e. the effect of the in-plane fields $H_{\parallel}$, as shown, on $\rho_{ab}(H_{\perp})$ at $T=0.019$~K and $T=0.315$~K, respectively.  Thin solid lines guide the eye.
}
\label{B-B1}
\end{figure}
%

\clearpage

\begin{figure}
\centerline{\includegraphics[width=0.68\textwidth]{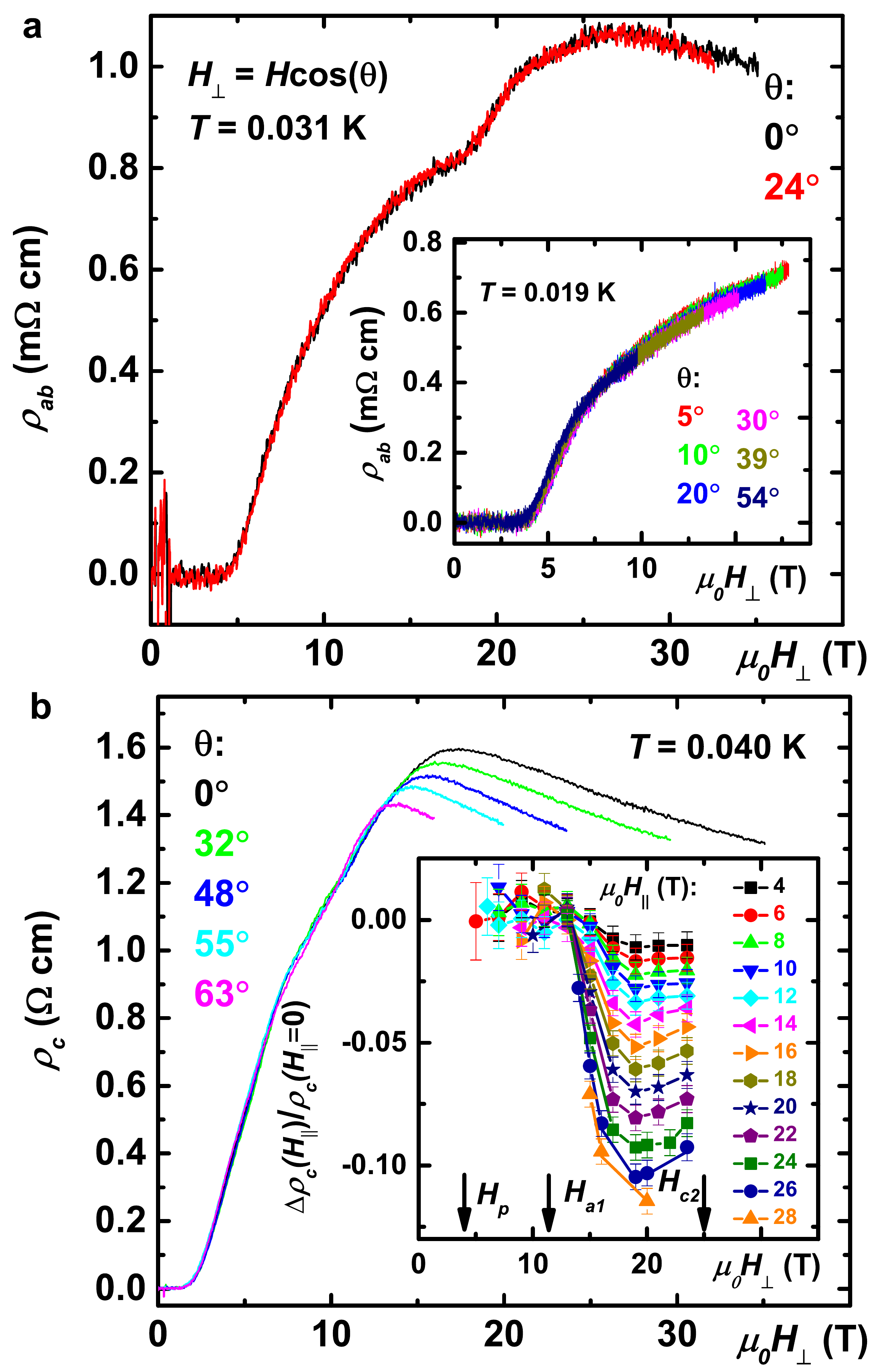}}
\caption{\textbf{Angle-dependent in-plane and out-of-plane resistivity of La$\bm{_{1.48}}$Nd$\bm{_{0.4}}$Sr$\bm{_{0.12}}$CuO$\bm{_4}$ vs $\bm{H_{\perp}=H\parallel c}$ axis at low $\bm{T}$, as shown.}  \textbf{a,} $\rho_{ab}$, and \textbf{b,} $\rho_c$, for different angles $\theta$, as shown; the uncertainty $\Delta\theta= 0.5^{\circ}$.  The in-plane $H_{\parallel}=H\sin\theta$ does not affect $\rho_{ab}$; here $H_{\parallel}$ was oriented parallel to the crystallographic [110] (or [1$\bar{1}$0]) axis.  On the other hand, $\rho_c$ is reduced by $H_{\parallel}$; here the field was parallel to the crystallographic $a$ (or $b$) axis.  The inset in \textbf{b} shows the corresponding $\Delta\rho_c(H_{\parallel})/\rho_c(H_{\parallel}=0)=\rho_c(H_{\parallel})/\rho_c(H_{\parallel}=0)-1$; $T=0.040$~K.
}
\label{LNSCO-angle}
\end{figure}
%

\clearpage

\noindent\textbf{Supplementary Information References}

\noindent [S1] Qin, Y., Vicente, C. L. \& Yoon, J.  Magnetically induced metallic phase in superconducting tantalum films. \textit{Phys. Rev. B} \textbf{73}, 100505(R) (2006).\\
\noindent [S2] Giamarchi, T. Disordered Elastic Media.  \textit{Encyclopedia of Complexity and Systems Science.} (Ed. R. A. Meyers, Springer, 2009).\\
\noindent [S3] H\"ucker, M.  Electronic interlayer coupling in the low-temperature tetragonal phase of La$_{1.79}$Eu$_{0.2}$Sr$_{0.001}$CuO$_4$.  \textit{Phys. Rev. B} \textbf{79}, 104523 (2009).\\
\noindent [S4] H\"ucker, M.  Structural aspects of materials with static stripe order.  \textit{Physica C} \textbf{481}, 3--14 (2012).\\
\noindent [S5] Chiba, K. \textit{et al.}  $^{139}$La-NMR study of spin-flop and spin structure in \lsco\, ($x\sim 1/8$).  \textit{J. Low Temp. Phys.} \textbf{117}, 479--483 (1999).\\
\noindent [S6] Yang, K.  Detection of striped superconductors using magnetic field modulated Josephson effect.  \textit{J. Supercond. Nov. Magn.} \textbf{26}, 2741--2742 (2013).

\end{document}